\documentclass[manuscript,screen]{acmart}

\usepackage{soul,xcolor}\usepackage{soul,xcolor}
\AtBeginDocument{%
  \providecommand\BibTeX{{%
    \normalfont B\kern-0.5em{\scshape i\kern-0.25em b}\kern-0.8em\TeX}}}

\usepackage{array}
\usepackage{amsmath}
\usepackage{pifont}
\usepackage{booktabs}
\newcolumntype{L}[1]{>{\raggedright\let\newline\\\arraybackslash\hspace{0pt}}m{#1}}
\newcolumntype{C}[1]{>{\centering\let\newline\\\arraybackslash\hspace{0pt}}m{#1}}
\usepackage{longtable}
\acmDOI{10.1145/3494519}

\acmBooktitle{ACM Transactions on Software Engineering and Methodology}

\setcopyright{none} 

\begin{document}

\setstcolor{red}

\title{Industry-academia research collaboration and knowledge co-creation: Patterns and anti-patterns}

\author{Dusica Marijan}
\email{dusica@simula.no}
\orcid{1234-5678-9012}
\affiliation{%
  \institution{Simula Research Laboratory}
  \city{Oslo}
  \country{Norway}
}

\author{Sagar Sen}
\email{sagar.sen@sintef.no}
\affiliation{%
  \institution{Sintef Digital and Sweetzpot}
  \city{Oslo}
  \country{Norway}}

\renewcommand{\shortauthors}{Marijan et al.}

\begin{abstract}
  Increasing the impact of software engineering research in the software industry and the society at large has long been a concern of high priority for the software engineering community. The problem of two cultures, research conducted in a vacuum (disconnected from the real world), or misaligned time horizons are just some of the many complex challenges standing in the way of successful industry-academia collaborations.
This paper reports on the experience of research collaboration and knowledge co-creation between industry and academia in software engineering as a way to bridge the research-practice collaboration gap. Our experience spans 14 years of collaboration between researchers in software engineering and the European and Norwegian software and IT industry. Using the participant observation and interview methods we have collected and afterwards analyzed an extensive record of qualitative data. Drawing upon the findings made and the experience gained, we provide a set of 14 patterns and 14 anti-patterns for industry-academia collaborations, aimed to support other researchers and practitioners in establishing and running research collaboration projects in software engineering.
\end{abstract}

\ccsdesc[500]{Software and its engineering~Collaboration in software development}

\keywords{Industry-academia collaboration, research collaboration, research co-creation, software engineering, technology transfer, knowledge transfer, collaboration gap, collaboration model, patterns, anti-patterns}

\maketitle

\section{Introduction}
The relevance of software engineering (SE) research and its potential to make a meaningful impact for industry practitioners has been an ongoing concern in the SE community. Bridging the industry-academia (IA) collaboration gap is a long-standing ambition of the SE community, recently given added attention by some of the prominent researchers in the field. For example, Briand acknowledges the limited impact of SE research on practice \cite{b1}, discussing some of the root causes for this situation, such as limited focus on real engineering problems, as well as a flawed reward system giving most credit to publication metrics. He also argues that there is a large disconnect between research and practice in SE \cite{b2}. Because research is not grounded in a real world setting, the output produced by research is not applicable or scalable. To overcome this problem, he advocates the case for context-driven research \cite{b3}, which means that research must focus on the problems driven by concrete needs in specific domains in order to be impactful. Similar to this view, Shneiderman \cite{b4} believes that a problem-oriented approach to research, which incorporates both theoretical developments and validated solutions ready for deployment, is the key to bridging the IA gap. Based on four decades of experience in the software industry and SE research, Selic \cite{b5} further claims that transferring the output of software research into useful industrial products often fails. If research and its productization are not intertwined processes, it can lead to failures of both the research and the productization.

Many existing approaches to implementing IA collaboration are based on some form of a technology or knowledge transfer process, such as the Gorschek's model for technology transfer in practice \cite{Gorschek}. A typical technology transfer process assumes that the research problem and its solution are (primarily) created by researchers and then transferred to practitioners. We, however, take a different stance on IA collaborations. We believe that research knowledge co-creation is a key to successful IA collaboration. We define co-creation as the process of participative value creation, where industry and academia actively participate in problem definition and solving, aiming to develop more relevant solutions for all participating parties, thus reducing the risk of the research collaboration failing. In this paper, we report on our 14-year experience of applying research knowledge co-creation to IA collaboration in SE.  
Our experience is derived from and validated in three different environments. The first is an 8-year research collaboration project between a research lab in SE and the Norwegian software industry. The second is a 5-year collaboration between a Norwegian start-up company and several international academic partners, conducting industry-relevant research. The third is an ongoing interdisciplinary collaboration in the context of a European H2020 project, involving 12 partners from across Europe.   
Using a combination of data collection and analysis methods we have collected and analyzed an extensive record of qualitative data on research knowledge co-creation. We synthesize this knowledge in the form of 28 patterns and anti-patterns, as a set of good and bad solutions to the challenges of IA collaboration in SE. Compared to existing work in this area, our patterns and anti-patterns are presented as reusable solutions describing the exact problem they address and its circumstances, a potential negative solution and how to recognize it and understand its implications, and a working solution and its benefits.

\section{Background}
IA collaborations present opportunities to yield benefits for both participating parties, such as increased access to new knowledge for industry and a validation context for scientific results for researchers \cite{b24}. However, a substantial challenge has been associated with building scalable and effective research collaborations between industry and academia in SE \cite{b6}. Next, we review some of these challenges, followed by lessons learned on IA collaboration.

\subsection{Challenges in Industry-Academia Collaborations}
Challenges in IA collaborations exist on both sides of the collaboration. Chimalakonda \cite{b7} notices the trend where most practitioners believe that researchers work on dated or futuristic theoretical challenges which are divorced from industrial practice, while researchers believe that practitioners are looking for quick fixes instead of using systematic methods. Researchers seem to be more interested in proposing new techniques and tools, focusing on technical novelty, while practitioners are interested in solutions that work in their context, regardless of novelty. Such views of researchers and practitioners often make the two seem like separate islands. On one side there are practitioners who do not see much benefit of research results and on the other side there is the "ivory-tower" of academia. Runeson \cite{b8} draws attention to the challenge of different time horizons of IA collaboration projects, which are generally shorter in industry compared to academia. This circumstance, as he observes, can cause friction and frustration on both sides of the collaboration. Bern \cite{b27} further argues that one of the reasons for the disconnect between research and practice lies in the difficulty for practitioners to make use of the knowledge produced by academic researchers. Academic knowledge is typically disseminated through scientific papers, which are difficult to absorb by many practitioners. He suggests that in addition to disseminating research results as scientific papers, researchers could use blogs, videos and interactive posters to engage and interact with practitioners. Ivanov \cite{b9} gives more examples of gaps between research and practice, indicating a wide disconnect between research topics discussed at academic conferences and the problems that practitioners are looking to solve.

A panel at ICSE 2011 titled "What industry wants from research" 
tried to identify key challenges and ways forward for bringing theory and practice closer together. Some of the panelists pointed out that potential reasons for the gap between the two lie in the constrained scalability of research solutions, limited interest of researchers for industry needs versus publications, and insufficient attention given to the integration of research into practice. To reduce the gap between industry and academia, suggestions were made for industry to tell researchers their real problems and share their case studies, and for researchers, to work on improving the usability of tools and methods, and to consider how to improve value, especially earning.

Furthermore, Wohlin \cite{b12} analyzes success factors for two IA collaboration projects in SE in Sweden and Australia, concluding that the industrial side of collaboration is the key element for successful collaboration. He also discusses top ten challenges for IA collaborations, some of which include trust and respect, roles and goals, and knowledge exchange instead of technology transfer, which we have also seen as important in our collaboration projects. In addition, he proposes five levels of closeness between industry and academia \cite{b13}, ranging from less to more close: not in touch, hearsay, sales pitch, offline, and one team. 

We have observed many of these (and some other) challenges in our IA collaboration experience. In the collection of patterns and anti-patterns presented in this paper, we discuss potential solutions to such challenges. In Table \ref{tab:long} we summarize which of our proposed solutions are relevant for these (and other) challenges reported in the context of IA collaboration projects.   

\subsection{Lessons Learned on Industry-Academia Collaboration}
Lessons learned from IA collaborative projects are often reported in literature. For example, Sandberg \cite{b15} reports lessons learned in a long term IA collaboration, describing success factors enabling research activities and results, including continuity, communication ability, industry goal alignment, and deployment impact. They target collaborative practice research (CPR), which is a collaboration practice relying on action research, experiments, and practice studies. While CPR is similar to the collaboration practices explored in our work, their experience involves only one company and an academic research institute, while our experience stems from a much larger context, involving 5 academic groups and 17 industry partners in total. Besides, while this study presents insightful success factors for IA collaboration projects, in order to apply the success factors, we see a need to systemically describe the context and problems that these success factors can be useful for, as described in our study.  
Dittrich \cite{b28} summarizes the experience of applying Cooperative Method Development (CMD) approach to IA research projects. In their experience, action research, as one of the applied principles, consists of three phases: understanding, deliberating change, and implementation and evaluation of improvement. While these three phases can be generally considered as part of the collaboration framework we applied in our collaboration contexts, we also focus on the importance of building the culture of one team consisting of participants from industry and academia, both equally participating in all stages of research knowledge creation, deployment and use, as one of the differences with CMD. While CMD aims to answer the questions of how software development practitioners tackle their everyday work, and how can methods, processes and tools be improved to address the problems experienced by practitioners, we, in addition, aim to uncover how to involve practitioners in the research knowledge generation process, as the final users of such knowledge. Marijan \cite{Marijan} reports the Certus collaboration model, consisting of 7 elements of IA collaboration that enable the culture of participative knowledge creation. These elements include problem scoping, knowledge conception, knowledge and technology development, knowledge and technology transfer, knowledge and technology exploitation, organizational adoption, and market research. While this model provides a systematic way of structuring different phases of collaboration, thus hopefully increasing its success, we further see a need to provide a catalog of patterns and anti-patterns that researchers and practitioners can use in a wide range of practical challenges occurring at different phases in collaboration. Barroca \cite{b29} presents lessons learned from three agile project with industry. These lessons include building trust, having written agreements in place, flexibility, providing outputs tailored to different audiences, cash investment from industry, having relevant research expertise, and regular contacts with industry. The concept of agile research targeted in their work is similar to our collaboration practices. However, their work reports overall lessons learned from applying agile research in three case studies, without special focus on explaining how specific collaboration challenges can be addressed by specific solutions. Mathiassen \cite{b30} summarizes experiences from an IA collaboration focusing on goals, approaches, and results. He underlines the importance of combining action research, experiments, and conventional practice to balance between relevance and rigor in IA collaborative research. While this work provides useful findings and lessons learned, these findings are lacking evaluation in a longer and larger context. Besides, no mapping is provided between lessons and specific collaboration challenges they address. Sjoo \cite{b16} applies a systematic literature review to identify general factors enabling collaborative innovation between industry and academia according to a project timeframe. These factors include resources, IPR, boundary-spanning functions, prior collaborative experience, culture, status (reputation), environmental factors (geographical proximity). Some of these factors, such as IPR, experience, culture have proved essential in our experience. Based on the number of projects describing IA collaborations in SE, Garousi \cite{b10,b11} synthesizes a comprehensive list of 63 collaboration challenges and 127 best practices applicable to these challenges. This is, to date, the most comprehensive review and a mapping between various reported challenges and their potential solutions. While we observe many of these 63 challenges in our collaboration experience, our work also presents several novel challenges. There are also few challenges summarized in \cite{b11} which we did not have in our experience (see Table \ref{tab:long}). In Table \ref{tab:long}, we provide an extensive comparison of our presented patterns and anti-patterns with best practices synthesized in \cite{b11}, relative to the set of challenges coming from both our experience and identified in \cite{b11}. From this table we can also see which challenges are unique in our experience, as well as solutions to particular challenges that differ from previously reported solutions for the same challenges. In another two studies by Garousi \cite{Garousi2016, Garousi2019}, the authors analyze 10 projects in two countries they participated in (\cite{Garousi2016}) and 101 external projects in 21 country using a survey method (\cite{Garousi2019}) to investigates relative impact of the collaboration challenges and related solutions, as reported in \cite{b11}, on the projects they study. Different from these studies \cite{b11, Garousi2016, Garousi2019}, our work presents a collection of reusable solutions, explaining when to use and when not to use a particular solution. The reusability is achieved by describing the problem and its causes, the context and circumstances under which the problem is applicable, the negative solution (solution that seems good but is actually ineffective), the effective solutions that works within the scope of the given context, the benefits and limitations of the solution, the consequences and related solutions (see figure \ref{fig_template}). In another study \cite{Garousi26}, Garousi reports seven lessons learned on how to conduct successful IA collaborations, based on the experience from 26 projects. The lessons are: working as one team (also recognized by \cite{b13}), identifying the research problem, ensuring practicality and applicability of approaches, conducting cost-benefit analysis of approaches, need for maturity of research prototype tools, encouraging further adoptions, long term benefits and benefits beyond the involved parties. While providing useful insights, in the description of the lessons the authors do not discuss what specific problems the lessons address, neither they mention alternative but ineffective ways of addressing the challenge, nor how can one recognize if the lesson should be applied, nor what are the benefits and consequences of using the lesson. What makes our study different from this and other existing work is a reusable presentation of solutions to IA collaboration challenges, which enables other researchers and practitioners to reuse the solutions in their context and avoid mistakes. Our findings originate from our own experience of IA collaboration projects, unlike in studies \cite{b11} and \cite{Garousi2019}, which report a literature review and survey findings. Furthermore, while \cite{Garousi2019} recognizes only an industrial impact created by solutions produced in the collaboration, we, in addition, give importance to an academic impact. In our experience, academic impact (in addition to industrial impact) proved to contribute to the commitment and sustainability of IA collaboration projects. 

\section{Research Knowledge Co-Creation}
\begin{figure*}
\centering
\includegraphics[trim=0cm 0cm 0cm 0cm, width=5.5in]{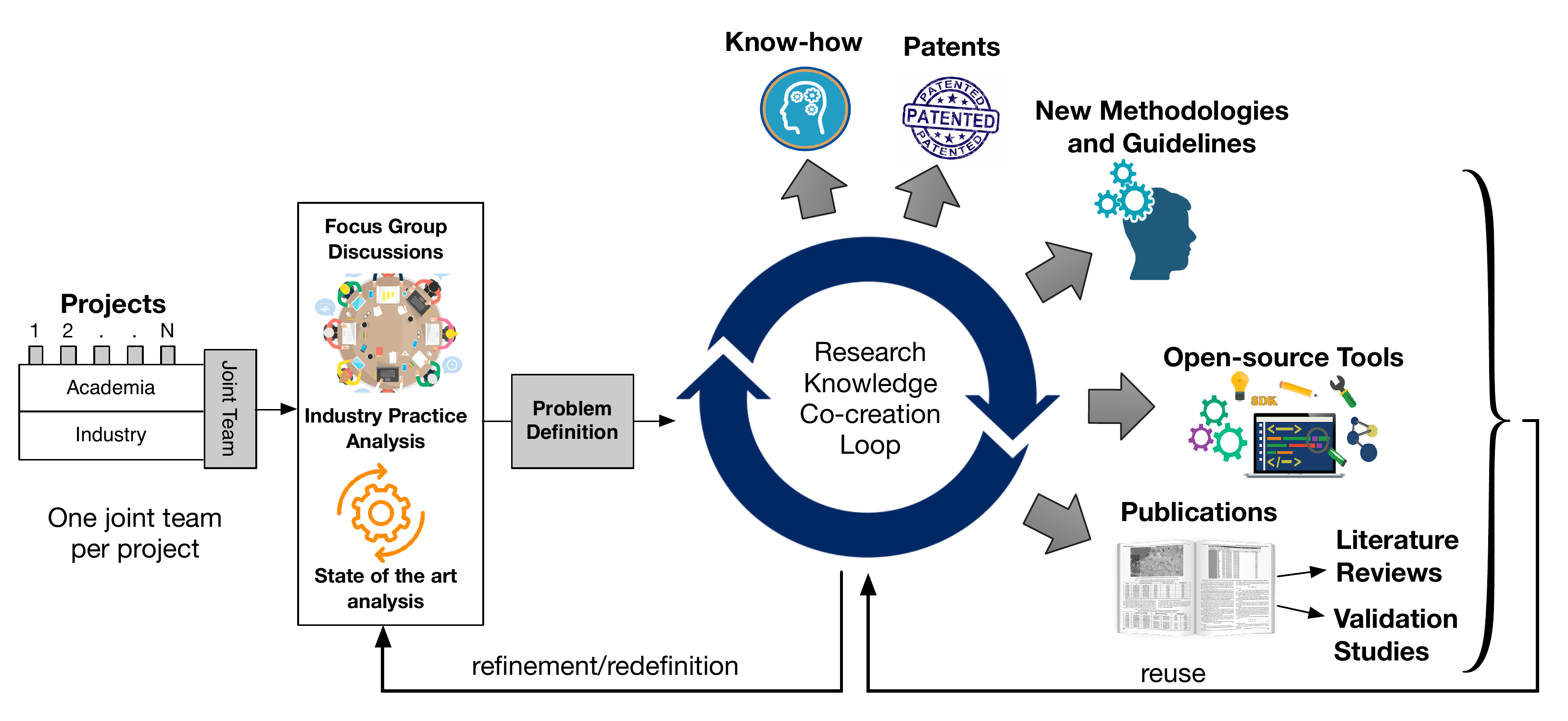}
\caption{Co-creation process.}
\label{fig_image}
\end{figure*}

IA collaboration projects often structure the collaboration practice using some form of a technology transfer process, such as \cite{b19,b31}. Technology transfer has been traditionally understood as "the process of transferring technology from the person or organization that owns or holds it to another person or organization" \footnote{Wikipedia.org, Technology transfer}. We believe that such a practice does not bring forth the best results in IA collaborations. This is because transferring technology "from" an organization "to" an organization depicts a disconnect between the two organizations, with little opportunities for aligning the goals of both parties for the solution being developed and creating a common ground with respect to the desired solution "-ilities" \cite{b5}, denoting scalability, usability, dependability, integrability, evolvability, maintainability, and security, to name a few. Instead, we argue that research knowledge co-creation, where industry and academia actively participate in all phases of the collaboration project, from problem definition to solution exploitation, creates the culture of co-ownership and stands the best chance of creating benefits for both parties. More specifically, we define \textit{research knowledge co-creation}, in the context of industry and academia collaboration in SE, as the process of collaboratively solving problems collaboratively defined by academic researchers and industry practitioners to generate both scientific knowledge and innovation in the form of new methodologies and guidelines, software libraries/tools, publications, and patents. The fundamental tenet in co-creation is that ideas are created and improved together instead of individually at each side.

The culture of co-creation is graphically presented in Figure 1. A joint team is created per project, consisting of academic researchers and industry practitioners, who work together through all the stages of co-creation. At beginning, intensive focus group discussions are held to thoroughly understand the industry context. The state-of-the-art is analyzed in relation to the observed industry practice, to identify any existing solution that can assist problem solving. Based on this knowledge, a joint problem is defined, which is being redefined as industry practice changes and as more knowledge is produced in the collaboration. The output of co-creation are new tools and methodologies, patents, know-how, co-authored academic publications, all of which, when applied to industry practice, can improve different aspects of it.

While the co-creation concept has been known in the business context for customer relationship building \cite{b25} \cite{b26}, it is not so used in the IA collaboration context. Our motivation for using co-creation to organize IA collaboration is to involve industry practitioners \textit{as the final users of research knowledge in the knowledge generation process}. As a result, they can own the created knowledge instead of having to invest extra effort to absorb the knowledge from academic publications, which are often in a form targeted towards researchers and not average practitioners \cite{b27}. Furthermore, involving industry practitioners as problem owners in the problem definition process helps ensure that the relevant/right problem is addressed and solved during the collaboration.

\subsection{Relation to Other Forms of IA Collaboration}
We contrast research knowledge co-creation, as considered in this paper, to action research, innovation and technology transfer.

\textit{Action research} \cite{Avison} is a well-known paradigm in the context of SE research for IA collaboration. Action research is intervention-focused, where an action is introduced by a researcher with the goal of understanding the consequence of the action in the studied context. The researcher also participates in implementing the action and making observations of the action. Action research is iterative, in the sense that the action is improved over multiple cycles. An analogy of action research in public health could be to administer a vaccine within a population and observe infection levels and health. The interventionist approach to research is the center piece of action research. One of the drawbacks of the interventionist mindset often very much depends on the researcher in terms of implementing the action and recording observations since, as the word suggests, it is an intervention with regards to the routine of a practitioner. Co-creation, on the other hand, does not start with an interventionist mindset. Instead, co-creation aims to generate commensurate value for all stakeholders in the team where the researcher works with a practitioner to create artifacts such as tools, publications, and patents. An intervention may also result in artifacts such as tools and publications, but the sponsoring thought behind the intervention is to intervene based on a predetermined hypothesis.

\textit{Innovation} is the practical implementation of ideas that result in the introduction of new goods or services, or improvements in offering goods or services \cite{Schumpeter}. Innovation is not tied to a particular approach to implementation such as co-creation. However, co-creation can very well be a means to innovate or create innovative goods and services. 

\textit{Technology transfer}, as the name suggests, takes results and prototypes conceived in a research institute or a university and attempts to commercialize them through various instruments, such as licensing, spin-offs, and patents, often followed by a market study also financed in part by the government. Technology transfer office (TTO) at research institutes and universities is responsible for technology transfer and other aspects of the commercialization of research. 
On the other hand, the starting point for co-creation is to define a problem together, whose solving may lead to a product, instead of aiming to commercialize a preconceived prototype or approach. The ownership structure in co-creation is very different compared to TTO where ideas from both research and practice, and commercial value of research results are debated in the very beginning. People involved in the co-creation process feel a sense of ownership and are given a fair chance to think about risk and resources at the time of conception. In the context of this paper, co-creation is focused on generating SE research knowledge in the form of patents, publications, software tools, services, standards etc. In the case of a TTO, there is often a lack of a driving force, as the inventor, who is a researcher, may not be business savvy and may expect someone to take the idea further into the market with only a peripheral and supervisory involvement and a hands-off approach. The reason for this entails the risk involved in taking a leave or quitting a stable academic position. Moreover, many research results from a university may not consider commercial value early in the process of conception. The TTO office has an administrative role in the project, with the goal of ensuring legal rights of ideas and inventions for the university or a research institute through selective investment into projects. Hence, TTOs are poorly suited to facilitate co-creation and foster mutually beneficial ownership structures between researchers and industry practitioners. A summary comparison of co-creation with the other three forms of IA collaboration is given in Table \ref{tab:comparison}.

\begin{table}[ht]
\centering
\caption{Comparison of different industry (I) and academia (A) collaboration methodologies.}
\begin{tabular}[t]{l>{\centering}p{0.12\linewidth}>{\centering}p{0.12\linewidth}>{\centering}p{0.12\linewidth}>{\centering\arraybackslash}p{0.12\linewidth}}
\toprule
&\textbf{Co-creation}&\textbf{Action research}&\textbf{Innovation}&\textbf{Technology transfer}\\
\midrule
Implies \textit{how-to} (the process of collaboration)& \checkmark & \checkmark &  & \\
I and A involved in all stages of collaboration&\checkmark&&Can be&\\
Based on reciprocity &\checkmark&&Can be&\\
Iterative&\checkmark & Can be& Can be&\\
Free of pre-determined hypothesis& \checkmark &  & Can be & \\
Knowledge creation and exploitation separated& & Can be & Can be & \checkmark \\
\bottomrule
\end{tabular}
\label{tab:comparison}
\end{table}%

\subsection{Output of Co-creation}
The output of co-creation is manifested in different forms, including knowledge, products, services, and methods. \textit{Co-created knowledge} is embodied as (a) know-how contributing to intangible benefits, such as learning and knowledge transfer, as well as increased innovation capacity of industry practitioners and the ability to absorb external scientific knowledge disseminated through academic publications, (b) publications with joint authorship between academia and industry, and (c) patents with some form of joint ownership between legal entities represented by industry and academia. 
\textit{Co-created products}, in the context of SE, refer to software libraries, tools, cloud-based services with documentation, released with a software license that both academia and industry agree to. For instance, a typical approach is to release an open-source community version of a software library or a tool and a closed-source customized version used and improved by industry practitioners. 
\textit{Co-created services} are consulting, educational, and certification services created by IA collaboration partners. Such services may include requirements elicitation to fit solutions developed in co-creation to new contexts emanating from within the existing partnership or from new partners from outside the consortium. For instance, a service could be to systematically adapt a software tool for testing highly-configurable systems, which was initially developed for video systems, to testing robotic systems. Similarly, services could involve creation of evidence in the form of validation studies to support certification. 
\textit{Co-created methods} are jointly developed systems or ways of doing something in the context of SE, such as requirements specification, software implementation, software architectural design, software evolution, and software validation and testing. The essence of co-creation here is to customize methods and their timeliness to the needs of both industry and academia.

\section{Collaboration Contexts}
\label{Contexts}
Our experience of research knowledge co-creation stems from three IA collaboration projects. One is a Large-Scale and Long-Term (C1: LS-LT) collaboration, the second is a Small-Scale and Mid-Term (C2: SS-MT) collaboration, and the third is a Large-Scale and Mid-Term (C3: LS-MT) collaboration, henceforth referred to as C1, C2, and C3. The differences and similarities between these three collaborations are described in Table \ref{tab:all}.

\begin{table}[ht]
\centering
\caption{Three studied collaboration contexts.}
\begin{tabular}[t]{l>{\raggedright}p{0.23\linewidth}>{\raggedright}p{0.23\linewidth}>{\raggedright\arraybackslash}p{0.23\linewidth}}
\toprule
&\textbf{C1: LS-LT}&\textbf{C2: SS-MT}&\textbf{C3: LS-MT}\\
\midrule
Duration&8 years&5 years&3.5 years\\
Number of partners& 8& 4&12\\
Domain& Software V\&V & IoT & Industry 4.0\\
Collaboration experience & Extensive& Little & Medium\\
Average team size&5 persons&5 persons&6 persons\\
Team skillset&Computer scientist, Software engineer, Software tester, Configuration engineer, Product manager & Computer scientist, Industrial designer, Software developer, Hardware engineer, Physicist
& Machine learning engineer, Data quality expert, Computer scientist, Software architect, Hardware engineer\\
\bottomrule
\end{tabular}
\label{tab:all}
\end{table}%

\subsection{A Large-scale Long-term Collaboration Context}
A large-scale long-term collaboration context is a Norwegian eight-year research-based innovation project that ran from 2011 to 2019. The project was hosted by Simula Research Laboratory (SRL), engaging a SE research group at SRL consisting of six senior researchers, six PhD students and one research engineer. On the industrial side, the collaboration involved four large companies in networking, robotics, and oil\&gas industry, one Small and Medium-sized Enterprise (SME), and two public sector institutions. Table \ref{tab:c1} summarizes the domains and key competences of all eight participating partners. Besides industry and academia, the project also involved the government, the Research Council of Norway (RCN), forming a Triple Helix system [22]. The project was funded by the RCN and an in-kind investment from our industrial partners. The motivation to involve in such a research-collaborative project for the industrial partners was the opportunity to get access to knowledge and resources they did not possess at that time, but which were essential for developing new and improving their existing software V\&V tools and frameworks. The motivation for the RCN to support such an IA collaborative project was to foster economic and social impact in Norway, by improving the software quality and thus competitiveness of the Norwegian software industry and public sector services. 

The main research direction of the collaboration project was software testing and V\&V, targeting: real-time embedded software systems, highly-configurable software systems, and data-intensive software systems. Research output in the project consists of scientific publications, methodology and process guidelines, and software prototype tools and services. The project was structured into 6 scientific sub-projects, each dealing with a specific thematic area of a broader field of software V\&V, and 3 governing sub-projects, dealing with overall project management and dissemination. 
Scientific sub-projects addressing V\&V problems common to several partners crosscut all partner domains where there is interest. In this way, we achieved an increased interaction and communication between the partners, relating to concrete problems and their technical solutions, which allowed for common areas of interest to arise and create synergies.

\begin{table}[ht]
\centering
\caption{C1: LS-LT Collaboration Context. Type: RI=Research institute, LI=Large industry, PS=Public sector service, SME=Small to medium enterprise.}\begin{tabular}[t]{p{0.25\linewidth}>{\raggedright}p{0.25\linewidth}>{\raggedright}p{0.05\linewidth}>{\raggedright\arraybackslash}p{0.4\linewidth}}
\toprule
\textbf{Partner}&\textbf{Domain and Competence}&\textbf{Type}&\textbf{Goals for Collaboration}\\
\midrule
SIMULA&Research on software V\&V & RI&Develop, validate, and deploy novel technologies for cost-effective software V\&V\\
CISCO & Video communication software& LI & Reduce regression test costs with test optimization\\
ABB & Robot control software & LI & Reduce the cost of test selection and scheduling\\
FMC TECHNOLOGIES & Subsea control software & LI & Reduce the costs of system (re)configuration\\
KONGSBERG MARITIME&Subsea control software&LI&Improve the cost-effectiveness of safety analysis and certification\\
NORWEGIAN CUSTOMS&Customs accounting&PS&Automate non-regression testing. Improve the quality of tests\\
ESITO&SW tools for domain driven development&SME&Create new market opportunities with research-based innovation\\
NORGES KRAFTREGISTERT&Cancer registry system&PS&Transform a manual system into an ICT-based system, while ensuring system quality\\
\bottomrule
\end{tabular}
\label{tab:c1}
\end{table}%

\subsection{A Small-scale Mid-term Collaboration Context}
Our small-scale mid-term collaboration context involves an Internet of Things (IoT) startup Sweetzpot (SZ) and several academic partners. SZ develops a sensor to monitor breathing patterns from ribcage and abdominal incursions and excursions. SZ also develops web and mobile applications to interface with sensors to receive raw breathing data and extract relevant information such as breathing rate, respiratory flow, and breathing patterns of interest, such as apneas. At the point of data collection, SZ had about 12 employees with very diverse skills, consisting of computer scientists, a physicist,
an industrial designer, software developers, a hardware engineer, along with people in sales and marketing. 
The partners in the collaboration are listed in Table \ref{tab:c2}, along with their competencies and interest areas.  

The main research direction for the collaboration was to create knowledge and insight from sensor data. The motivation for SZ to involve in collaborative research projects with academic teams stems from the creation of mutual value in the form of a) scientific validation of the SZ's technology in diverse scenarios, b) creation of open-source software for community building around the SZ's technology, and c) development and bi-directional transfer of skills between a startup and academic teams. 
The motivation for academia to participate in the collaboration with a startup is supported by various funding instruments and encouraged by funding agencies. These collaborations create skills necessary for improving the innovation capacity of companies and further the economic development of countries. 

The scientific output of the collaboration are open-source code repositories, mobile apps for scientific data collection, mutual development of skills, and several scientific publications. The collaboration between some of the academic groups happened when results such as open source software or data quality studies from one group were of relevance to another group. Data collection in this collaboration context was performed during a 5-year period, from 2015 to 2020.

\begin{table}[ht]
\centering
\caption{C2: SS-MT Collaboration Context. Type: SME=Small to medium enterprise, RI=Research institute, U=University.}
\begin{tabular}[t]{l>{\raggedright}p{0.25\linewidth}>{\raggedright}p{0.05\linewidth}>{\raggedright\arraybackslash}p{0.4\linewidth}}
\toprule
\textbf{Partner}&\textbf{Domain and Competence}&\textbf{Type}&\textbf{Goals for Collaboration}\\
\midrule
SWEETZPOT&Sensor hardware and software
&SME&Develop and test breathing sensor hardware and software applications for research\\
RITMO CENTER & Research on time and motion & RI & Research breathing rhythm during musical performance and listening\\
UNIVERSITY OF OSLO & Research on sleep apnea & U & Research detection of obstructive sleep apnea with low-cost sensors\\
UNIVERSITY OF LEEDS & Research on pollution & U & Research effect of pollution on respiratory flow\\
\bottomrule
\end{tabular}
\label{tab:c2}
\end{table}%

\subsection{A Large-scale Mid-term Collaboration Context}
Our large-scale mid-term collaboration context is an interdisciplinary 40-month European project entitled Interlinked Process, Product and Data Quality framework for Zero-Defects Manufacturing (InterQ) that had a kick-off in 2020 during the COVID-19 pandemic. In this article, we discuss the collaborative context in a specific work package in the project called \emph{InterQ-Data} that is concerned with improving data quality in Industry4.0. The collaborative context involved partners from Spain, Italy, Norway, Greece, France, and Germany. The objective of the collaboration is to perform different tasks on sensor data obtained, such as in-motion data quality validation, historical data quality validation, erroneous data repair, data quality as a service, while observing a manufacturing process. Sensor data is acquired from CNC milling, turning, and broaching machines in addition to sensors in harsh environments, such as thermo-couples in machine tool tips and acoustic emission sensors on a noisy shop-floor. The data acquired from sensors at sub-microsecond sampling periods can have missing data due to connection losses, can be duplicated (a.k.a. time collision), and can be outside the operating range of a sensor due to high temperatures. The project needs to ensure high accuracy and completeness of the sensor data, as well as to perform erroneous data repair when low quality data is detected. 

The project consists of both academic and industrial partners with complementary roles, with competence in software engineering, machine learning, AI, data quality management and advanced manufacturing. Table \ref{tab:c3} summarizes the domains and key competences of all 10 participating partners. 

One of the hallmarks of this collaboration is the fact that it has been forced to be remote and where all communication has been organized using a communication platform called Microsoft Teams due to the COVID-19 pandemic. This implies that partners are not physically meeting each other and visiting industrial sites.

\begin{table}[ht]
\centering
\caption{C3: LS-MT Collaboration Context. Type: MI=Medium-size industry, LI=Large industry, RI=Research institute, U=University.}
\begin{tabular}[t]{l>{\raggedright}p{0.3\linewidth}>{\raggedright}p{0.05\linewidth}>{\raggedright\arraybackslash}p{0.4\linewidth}}
\toprule
\textbf{Partner}&\textbf{Domain and Competence}&\textbf{Type}&\textbf{Goals for Collaboration}\\
\midrule
IDEKO S COOP&Advanced manufacturing&MI&In-motion data validation from machine tools\\
DNV GL AS & Data quality as a service & LI & Develop data quality as a service\\
SINTEF & IoT and AI systems & RI & Virtual sensing for erroneous data repair\\
RENAULT ESPANA SA & Cylinder heads for electronic cars & LI & Provide data from sensors during head milling\\
COMAU FRANCE SAS & Machine service provider & LI & Data acquisition from CNC machines\\
DANOBAT & Manufacturer of machine tools & LI & Provide machine tools for manufacturing\\
INLECOM & Digitization technologies & MI & Anomaly detection in sensor data\\
FUNDACION TEKNIKER & Advanced manufacturing research & ML & Achieve data quality in the energy sector\\
TU DARMSTADT & Data science in production & U & Uncertainty estimation\\
PREDICT SAS & Predictive maintenance in machining & MI & Temporal clustering of historical data from machining processes\\
\bottomrule
\end{tabular}
\label{tab:c3}
\end{table}%

\section{Research Method}
In this paper, we are interested in answering the question \textit{"What practices to follow and what to avoid to enable successful IA collaboration in SE?"} Given a certain IA collaboration challenge (e.g. industry practitioners not committed to collaboration), we are interested to know what practices are useful in avoiding or mitigating that challenge.

To answer this research question, we applied the participant observation and interview methods to collect a comprehensive record of qualitative data pertaining to three co-creation-based research collaborations spanning 14 years in total. We analyzed and correlated the collected data using a grounded theory approach, and observed that many of the findings are consistent across multiple projects. Consequently, we synthesized 28 patterns and anti-patterns for guiding the process of building and running successful IA collaborative projects in SE. 

\subsection{Data Collection}
\label{Collection}
Our goal for data collection was to capture both positive and negative experience of IA collaboration, and to find out what worked and what did not, and why, in different collaboration settings. We used a combination of research methods and data sources to collect qualitative data for our study. The primary data source were field notes from participant observation and transcripts from interviews. Other data sources included emails exchanged between the two observers and other project members on the topic of IA collaboration. Participant observation is a method of a systematic and unobtrusive data collection through social interaction between an observant and informants \cite{b32,b33,b34}, while an interview is a method of listening and talking to people to gain knowledge from individuals \cite{b35}. Throughout the process of data collection, as more data would become available from different sources, we would apply data triangulation, combining evidence coming from different sources, to increase the precision of our empirical study.

Data collection took place is cycles, interwoven with the data analysis cycles. In the first phase, we observed collaboration contexts C1 and C2. The observations took place during a number of occasions, such as project meetings, workshops and focus group discussions, and daily technical interactions among project members. As the two observers, who are also the authors of this paper, were active researchers in the two collaboration projects, they observed at least 90\% of all meetings, workshops, focus group discussions, and technical interactions (the other 10\% corresponds to the time they were off from work). The informants consisted of industry practitioners and academic researchers. The size of the group under observation varied from two to seven informants for project meetings, and up to 30 informants for workshops. The informants' background was in software engineering, product management, hardware design, scientific research, data science, computer science, and machine learning. The average work experience of each informant in their respective areas of expertize was nine years. 
Note taking during observation was performed as unobtrusively as possible, with an observer acting as a "normal" participant", to not make informants feel observed. The notes taken included diverse information including interaction patterns between the participants, expectations from the collaboration, terminology, decisions about technical choices, and frequency of status updates. Immediately after each observing session, the observers would read the notes and subsequently augment them with more details and reflections of the observer. After the first cycle of data collection, we analyzed the data using the process described in Section \ref{Analysis}. Next, we continued with another cycle of data collection using the semi-structured interview method. Specifically, the two observers interviewed five participants from each collaboration context, C1 and C2. The interviewees were selected purposively, based on the observer's judgement of which participants will provide the richest information relative to the questions that arose during the previous cycle of data analysis. In some data collection cycles, we also used a maximum variation sampling strategy, to ensure a wide range of different backgrounds and expertize. Industry practitioners and academic researchers were equally represented among the selected interviewees, consisting of software engineers, software testers, machine learning engineers, data scientists, product managers, researchers. The average work experience of each interviewee in their respective areas of expertize was seven years. We performed semi-structured interviews discussing specific questions pertaining to IA collaboration challenges and desired ways of working, but also probing deeper into IA collaboration aspects that the respondents seemed to have strong opinion about. Some examples of the interview questions are \textit{"What was the most critical challenge for the project?", "What solution, if any, worked for that challenge?", "What were the benefits of that solution?", "What were the limitations of that solution?", "Was there any solution you tried but it did not work?", "What is the most important learning on IA collaboration from the project?"} Immediately after each interview, the observers would make interview transcripts. After the second cycle of data collection, we analyzed the data again, where newly collected data was providing additional information needed for a full understanding of good and bad IA collaboration practices we set out to study. The cycles of data collection and analysis continued interweaving until newly collected data started yielding redundant information. The observers could clearly see the patterns and different categories of findings, which is when we started approaching saturation. At that point, we applied the member checking technique \cite{Lincoln}, presenting the findings to 15 stakeholders from the two studied collaboration contexts, where the goal was to get their opinion on the validity of the findings. They confirmed the validity of the findings, confirming that data saturation has been reached. In total, we performed six cycles of data collection, three of which used participant observation, and other three using semi-structured interviews. The process yielded 10 patterns and 14 anti-patterns. In the second phase, we observed the collaboration context C3, replicating the process of data collection and analysis described for the first phase. Specifically, after each data collection cycle, we would augment the corpus previously collected from C1 and C2, and then go back and forth between the data points of the entire corpus, comparing them to find interesting findings. The goal of this process is to ensure that both already-defined and newly-defined patterns and anti-patterns hold for all collaboration contexts. After saturation has been reached (i.e. new data stopped adding new insights to the patterns), we augmented the set of previously defined patterns with additional four (P11-P14), which also held for C1 and C2 contexts. Finally, we applied member checking with stakeholders of all three collaboration contexts, confirming the validity of the complete set of 14 patterns and 14 anti-patterns. In total, we interviewed 45 interviewees, and observed 90 informants.

\subsection{Data analysis}
\label{Analysis}
For data analysis, we used an approach motivated by grounded theory, where the two observers first manually converted field notes from the observations and transcripts from the interviews into excerpts using open coding. Next, we compared and contrasted the excerpts with each other, looking for sets that relate to the same concepts. For example, the following two excerpts looked similar: \textit{"When we explain our daily practice to our academic partner, it is very difficult to use the same language they do, so they can understand us"}, and \textit{"Researchers often complicate things with their scientific language"}. We grouped together such sets of excerpts (in the example above, forming a code "practitioners have difficulty communicating with academia"). After we have built a set of codes, we started comparing them one to another. For example, the following two codes "practitioners have difficulty communicating with academia" and "academics have difficulty communicating with practitioners" came under a category named "difficulties in communication" (later translated into the AP6 name \textit{The incomplete tower of Babel}). The codes and categories were built by the observer who collected the data (one for each collaboration context), and then reviewed by the other observer, to prevent bias and reduce the subjectivity of coding. In case of dilemmas and opposing opinions, which did not happen, we would involve another project member to resolve the tie. Each cycle of data analysis was followed by a new cycle of data collection, as described in Section \ref{Collection}, after which we analyzed more excerpts and compared them with codes and categories to see whether they contradict, expand or support them (the latter denoted saturation). At the end, we had 28 distinct categories, which we present next.

\section{Patterns for Industry-academia Collaboration}
\begin{figure*}
\centering
\label{template}
\includegraphics[trim=-2cm 10cm 15cm 0cm, width=5in]{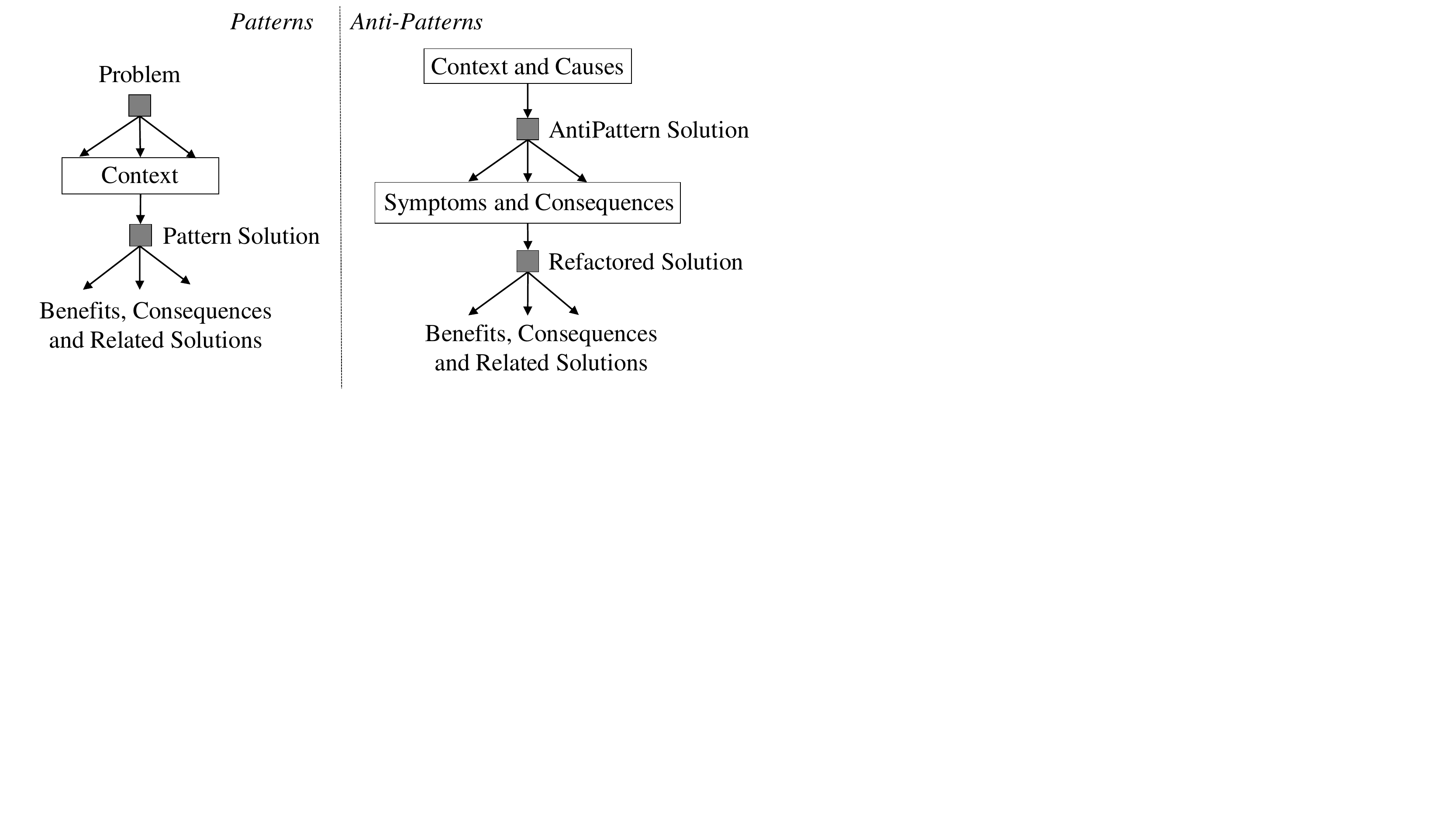}
\caption{Patterns and Anti-Patterns.}
\label{fig_template}
\end{figure*}

Patterns are reusable solutions to commonly occurring problems. In SE, they are often used in the context of design patterns \cite{b37}. Patterns encapsulate best practices that can be easily reused or adapted for solving recurring problems in a variety of situations. To support reusability, patterns are specified using templates. We use the following template for describing patterns: \textit{\textbf{Problem \ding{169} Context \ding{169} Pattern Solution \ding{169} Benefits and Consequences \ding{169} Related Solutions}}, illustrated in \ref{fig_template}. A pattern relates a particular \textit{problem} (in IA collaboration). \textit{Context} describes the circumstances under which the problem occurs or is applicable. A \textit{solution} describes how the problem can be resolved or eased in the scope of the context. The solution provides \textit{benefits}, but also may have remaining unresolved issues called \textit{consequences}. Finally, the solution can be \textit{related} to other solutions.  

\subsection{P1: Reciprocity Between Stakeholders}
\textit{Problem:} Stakeholders do not reciprocate in terms of effort.

\textit{Context:} Research collaboration process often requires investment of time, ideas, and effort from different stakeholders, to make the collaboration mutually beneficial. We quote
author James Baldwin who said, \textit{"Allegiance, after all, has to work two ways; and one can grow weary of an allegiance which is not reciprocal"}, as well as a researcher in C1 who said \textit{"It is so frustrating when our industry partner does not prepare what we asked for in the last meeting"}. Reciprocation can be hindered due to many factors such as lack in clarity of tasks, lack of resources, or external factors that take away focus from the project.

\textit{Solution:} To develop a norm for reciprocity between stakeholders, encourage an open and unambiguous communication. The give and take [25] psychology can help construct a reciprocal relationship. Identify the motivations and needs of all participants. Promote active dialog and frequent sharing of progress. It is important to acknowledge that not all project members are able to contribute equally to all project phases. The goal is to encourage a regular and well balanced exchange of knowledge where everyone contributes to the best of their capacity with some knowledge that other members of the project need. Reciprocity is often observed when the stakeholders reach a tipping point in collaboration. The tipping point of mutual trust is established when the joint activity is clearly linked to a well-defined and manageable goals that stakeholders manage to realize. These goals may include a publication to a specific conference or journal with a deadline, or it can be an application for project funding with a deadline, and it can also be a demo of a software tool to a third party at an event such as a conference or a trade fair. The time boundedness and the concreteness of the activity greatly helps establish reciprocity. Reciprocity can be measured in terms of commitment to tasks and attendance to meetings, co-authorship of research results (e.g. in terms of the number of words written by a stakeholder), time spent discussing a problem by all stakeholders. Our approach to address the reciprocity problem is to have, on the one hand, researchers who provide research artifacts that simplify daily routine of practitioners, and on the other hand, practitioners who can then effortlessly provide useful feedback for researchers on the practical usefulness of their research results. For instance, in C1, researchers provided a tool Depict \cite{b44} that Tool Customs used to verify customs declarations in their daily work. \textit{"Depict seamlessly connected to our test database and automatically generated reports that made it easy for us to present what we have been working on with minimal effort. This allowed us to keep focus on our priority task, while integrating research ideas to simplify our routine activities."} SRL's strategy was to build tools that could \emph{minimize effort} from practitioners such that they could seamlessly use a tool in a matter of minutes to produce a useful contribution that could be presented and used for reasoning. In C2, reciprocity was an issue for the partner with fewer resources. SZ had several customers with diverging requirements but only few personnel to handle the customer needs. For a commercial business, the extra hours required to reciprocate requests from stakeholders, such as academia, had to be compensated as a paid consulting activity or by involvement in national and European projects to receive partial public funding. SZ also took the role of a sensor vendor with third-party support in many of these publicly funded projects. In C3, reciprocity between stakeholder is built into tasks where partners are required to co-create together, often even without prior joint work experience. For instance, advanced manufacturing companies are required to acquire high velocity sensor data from the machining process and provide it to scientists working on computing data quality hallmarks and performing erroneous data repair for missing values.

\textit{Benefits:} Mutual learning based on active exchange of knowledge, which creates more meaningful collaborations, increases enthusiasm, creates synergies, and helps the project progress faster. 

\textit{Related solutions:} P8, AP4.

\subsection{P2: Standardized Data Exchange}
\textit{Problem:} Limited or cumbersome sharing of potentially reusable artifacts.

\textit{Context:} Participants in a co-creation project dealing with related problems often have the possibility of exchanging (and reusing) datasets, algorithms, code repositories, and tools. However, this seldom happens because these artifacts are not easily interfaced with each other.   

\textit{Solution:} These artifacts can be seen as services. It is worthwhile implementing standard protocols for interfacing between these services, so to avoid reinventing the wheel. Seeing artifacts as services implies that an artifact is associated with a standard for information exchange that allows interaction with a larger ecosystem. In the case of a dataset, such an associated standard could be a database technology based on SQL standards, such as ISO/IEC 9075:2016\footnote{https://www.iso.org/standard/63555.html}. When the dataset is stored in RDBMS supporting the SQL 2016 standard\footnote{https://www.iso.org/standard/63555.html}, then SQL can be used to read, write and modify the data. Similarly, algorithms can be encapsulated as a RESTful webservice where functions to get/set variables and perform some form of computation obey the REST constraints. RESTful web APIs are typically based on Hypertext Transfer Protocol (HTTP) methods to access resources via URL-encoded parameters and the use of JSON or XML to transmit data. Stakeholders from industry and academia adhere to such standards to ensure their data and algorithms can be plugged into a larger ecosystem with minimal effort, increasing business opportunities. However, concerns with intellectual property rights and making artifacts easily understandable and usable may hinder the flow of artifact exchange across stakeholders. In C1, SRL developed tools that adhered to industry standards, along with simple and user-friendly documentation, to help stakeholders use artifacts in less than 30 minutes. The tool Depict \cite{b44}, for instance, could be deployed by Toll Customs in a matter of minutes on their large test and production databases. Depict could interface to several standard relational database management systems (RDBMS) such as Sybase, Oracle, PostgreSQL, MySQL to verify certain properties in the data. This allowed Toll Customs to easily transition from Sybase to Oracle during their migration. In C2, the use of standard protocols greatly facilitated interfacing of SZ's sensors with many other devices. For instance, the standardization activity between SZ and RITMO was based on the Open Sound Control (OSC) protocol. This created the possibility for RITMO to interface the SZ breathing sensor to all other sensors and actuators in their laboratory and simultaneously gather a large amount of research data. In C3, there is no standard interface to connect tools developed by computer scientists in the project and data generated by machine tool developers. However, streaming data from a machine tool is available 24/7 from a cloud service in JSON format for a specified input time period. The machine tools often work around the clock to manufacture parts. Computer scientists have access to this data stream through a REST API. All computer scientists that perform tasks such as anomaly detection, erroneous data repair, and data profiling have developed a common understanding of the type of data they can develop their tools on based on the REST API. Docker containers are used as a common approach to build tools that can be exchanged and used by partners without special installation requirements and deployed on both edge devices and the cloud.  

\textit{Benefits:} Greatly increased the ease of reuse between different projects and the possibility to benefit from other participants' experience. A software engineer participant in C3 pointed out that \textit{"The use of standards to exchange data combinatorially increases the possibilities in terms of how collaboration can occur between multiple data providers"}. Generic tools built based on standard data exchange formats can be downloaded and used by projects around the world. This can increase the impact of tools from scientific research and help secure both publicly and privately funded projects.

\textit{Related solutions:} AP2.

\subsection{P3: Quantifying Impact}
\textit{Problem:} Difficulties of measuring diverse impacts of an IA collaboration. Typical impact metrics include a publication and patent count. However these metrics do not capture a multifaceted impact of IA collaborations.  

\textit{Context:} In a simple interpretation, making an impact in IA collaboration means, for researchers, changing the current state-of-the-art for the better, and for practitioners, changing the current state-of-the-practice for the better. There is also a necessity to go beyond organizational impact and quantify how collaborations can have an impact on one or more of UN's 17 goals on sustainable development\footnote{https://sdgs.un.org/goals}. Accurately measuring the factors contributing to that change is challenging \cite{b38}, yet important because as Peter Drucker said "if you cannot measure it, you cannot improve it". 
 
\textit{Solution:} Measuring impact in IA co-creation requires public recognition of \emph{reciprocal actions} resulting in the creation of tangible and intangible assets that benefit the industry and society. Tangible assets can be new or improved technology, such as open source software, algorithms, methods, tools, process guidelines, and scientific publications. While intangible assets are innovation capacity, human relations, shared management, visibility and brand value of the collaboration. For tangible assets, impact can be determined by measuring return of investment (RoI) or other quantitative metrics, for example, economic benefits. For intangible assets, impact could be gauged based on the timeliness, completeness and follow-up, consistency, relative quality, and complementarity of reciprocal actions for a successful collaboration. The latter can often lead to breakthroughs in science when partners involved are from diverse domains and are able to make small sacrifices to explore uncertain problem domains. In C1, we measured an impact by computing the amount of time saved in completing a testing activities in a company. In C2, one way to measure impact can be to look at the diversity and extent of the use of the Flow sensor developed by SZ. The sensor found application in health, sports, and music. The sensor was used by the University of Oslo as a low cost alternative to predict obstructive sleep apnea \cite{b48}. 
A future impact of the technology can be quantified in the form of percentage of the target market of solutions for obstructive sleep apnea captured by the SZ sensor. In C3, the impact of improved data quality is quantified in the form of improved auditability of data quality in manufacturing and fostering a quality culture. Auditability is measured as a traceability of data quality hallmarks or statistical properties of sensor data quality, observing the manufacturing process. Quality culture is indirectly measured through the reduction of scrap rate in manufacturing. Maintaining a trace of data quality is valuable feedback to the entire manufacturing ecosystem, helping improve performance of operators and engineers. Thinking about societal impact, sustainable development is a principal factor in governance, and the impact of the co-creation process needs to be in alignment with UN's sustainable development goals \cite{b45}. These goals include no poverty, zero hunger, good health and well-being, quality education, gender equality, clean water and sanitation, affordable and clean energy, decent work and economic growth, industry, innovation, and infrastructure, reduced inequalities, sustainable cities and communities, responsible consumption and production, climate action, life below water, life on land, peace, justice, strong institutions, and partnerships. In C1, we collaborated on the improvement of video conferencing systems developed by Cisco systems. This effort is indirectly connected to the minimization of the need for air travel (number of trips avoided) and consequently addressing UN' goal of responsible consumption. In C2, a project manager at SZ said \textit{"We always adhere to responsible production aiming to minimize defects, which consequently minimizes waste in the production of our Flow sensors. This was achieved by keeping a close eye on the inventory (number of sensors) and using low cost production technology such as a pick-and-place machine for local production of electronic printed circuit boards."}

\textit{Benefits:} Calculating quantitative and qualitative metrics corresponding to different types of impact (value), tangible and intangible, enables more comprehensive impact measurement and effort recognition. It is often an intellectual exercise to connect our seemingly negligible local contributions to a larger goal such as sustainability. It is increasingly requested by both public funding bodies and investors to quantify and present such an impact as part of their work.

\textit{Related solutions:} AP13.

\subsection{P4: Active Dialog}
\textit{Problem:} Goal misalignment between industry and academia, as well as limited transparency regarding intermediate results produced at different stages of collaboration ultimately lead to a lacking interest and disengagement. 

\textit{Context:} IA collaborations often suffer from inadequate level of communication, both between and within industrial and academic teams \cite{b39,b40,b41} on different aspects including goals, expectations, results, time frames, and responsibilities. As industry and academia are generally two culturally different environments, insufficient communication between the two leads to the lack of commitment and further deepens the gap between them. 

\textit{Solution:} Promote active dialog between industry and academia at all stages of the collaboration, from problem understanding and definition to solution validation. Active dialog implies words and actions occur in close succession or concurrently. Regular dialog does not necessarily mean words acted upon. Hence, the emphasis on active dialog. Such a practice helps build a trusting environment in which a participative and mutually beneficial value creation can be implemented. Promote active dialog across academic teams, as well as across industry partners, which will lead to a higher level of reuse of research results. Active dialog can be promoted by providing different channels of interaction, such as physical and virtual collaboration and meeting space, as well as different communication and knowledge exchange tools. In C1, SRL created several platforms to maintain active dialog. Bi-annual partner workshops were organized at a conference location to assemble all stakeholders in one place for a day and half of discussions to gain clarity and define problems that are both interesting for researchers and useful for industrial partners. Long term projects can get weary and require constant re-invigoration. Therefore, SRL invited several external keynote speakers, and organized hands-on workshops to invigorate stakeholders with new ideas and increased clarity. The keynotes also helped keep the stakeholders updated on the state of the art. In C2, the projects were of shorter duration, leading to intense bursts of joint work. However, it was essential for SZ to keep stakeholders in the loop and build a network between customers for sharing of ideas and artifacts, and also for the establishment of new projects. In C3, we have an additional challenge of remote work, due to COVID-19, where active dialog is maintained by weekly meetings over Microsoft Teams and through close follow-up on research ideas and subsequent action points in these meetings. Due to a large size of the project, there is a hierarchical communication structure in the project where smaller working groups work on specific topics and the results are summarized and communicated upwards in a leaders' meeting. The different working groups also help manage potential conflicts of interest between competing companies.

\textit{Benefits:} Continuous goal alignment and progress update between the collaborating partners, which leads to the shared ownership of both the problem and the solution, which is further crucial to prevent the drop of interest and commitment. A researcher in C2 pointed out \textit{"I was amazed to see how quickly we converged on which feature should be developed first, once we started talking details with the engineers."} Maintaining active dialog and keeping people in the loop is a continual process. However, there is a thin line between annoying stakeholders with unnecessary emails and actually engaging in a communication that is mutually beneficial. It is necessary to adjust the volume and the velocity of the communication to maintain a healthy active dialog.

\textit{Related solutions:} P7, P14, AP6, AP14.
 
\subsection{P5: Power of synergy}
\textit{Problem:} Confined research value creation due to a dissociate efforts made by industry and practice separately.

\textit{Context:} The growing complexity of industrial software systems developed by the majority of our partners requires a combination of knowledge, skills, and efforts to develop cost-effective solutions for software engineering and testing. Such combination of knowledge and skills is hardly found in any individual or a team whose members are of the same background. Moreover, co-creating with experts from different domains can push engineering and validation of software systems to new horizons. Still, it happens that IA collaboration projects lack collective value creation, as each side of collaboration works in isolation. 

\textit{Solution:} Enact a \textit{managed} research value co-creation, which entails a project plan with deliverables and milestones, and a project timeline, analysis of risks and how to mitigate them. A project leader conducts regular meetings with all stakeholders to hear about updates and find synergy during co-creation. Partners have regular internal meetings with individual stakeholders in order to meet the deadline and develop satisfactory results. Activities need to be performed consistently, ideally avoiding a last minute dash. All meeting notes and decisions are documented for easier onboarding of new members. In such a managed co-creation environment, individuals with different background, expertise and perspectives are brought together to share knowledge and work on a joint problem whose solving brings benefits to all participating parties. Consequently, there is a great potential for the identification of additional common areas of interest and synergies between the partners, which can lead to breakthroughs in research and practice. For instance, in C1, the testing of high-end video conferencing systems developed by Cisco Systems with several variable parameters and requirements for high video throughput pushed software testing research at SRL to handle a very large combinatorial problem. This led to novel algorithms published as scientific articles \cite{MarijanMP, Marijan2013, Marijan2016, MarijanDevOps, MarijanOB, MarijanPSR} and software. In C2, SZ's Flow sensor was used as a wearable in health (obstructive sleep apnea), sports (force in a rowing oar), and music (breathing rhythms while performing). All these application domains pushed the scientists, hardware and software developers in the collaboration to improve many aspects of the product: increasing battery life through low energy consumption needed for overnight recordings, long distance Bluetooth communication and re-connection logic due to uncertain connectivity, and machine learning models to predict variables of interest such as respiratory flow and apnea events from raw sensor data. These innovative activities were not foreseen during the conception phase of the project. As a hardware engineer at SZ mentioned \textit{"I would have never thought that we will get this far with all these new features we developed for Flow sensor. It's because we all worked together bringing ideas from so many different fields."} In C3, one example of synergy is when computer scientists and electronic engineers co-create to enable data quality management at different levels of abstraction (real-time and aggregated data on the cloud) and across different domains of expertise (signal processing and machine learning). For example, electrical engineers from IDEKO manage data quality during analog to digital conversion in a real-time controller, by data conversion and aggregation. 
Computer scientists use this aggregated data from the real-time controller to perform data profiling. They use machine learning models to create virtual sensors, which are used to repair erroneous or missing data in one sensor based on data available from other sensors.

\textit{Benefits:} Collective creativity leads to more effective problem solving, often resulting in unexpected and non-obvious ways of value creation. The synergistic influences from different domains can make a tool or a product far more robust and appealing to many users.

\textit{Related solutions:} P1.

\subsection{P6: Minimum Viable Tools}
\textit{Problem:} Lack of objective evidence showing the practical value of the knowledge created in early stages of a collaborative project.   
 
\textit{Context:} Practitioners engage in research collaborations with the objective of applying in their context the knowledge created in the collaboration and relevant to their problems, for solving practical problems. Researchers on the other hand are typically concerned with creating generalizable knowledge, and it may take a while before first results of contextualizing such a knowledge are available. Meanwhile, practitioners are kept in dark, which makes them not invested in the collaboration effort, hampering long-term success. As a software engineering practitioner in C1 described \textit{"Every time researchers present an update, they explain some new scientific stuff, but I never seem to quite get what are they talking about"}.

\textit{Solution:} Develop a minimum viable proof of concept (MVPC) and tools able to demonstrate practical benefits of even simple research concepts in early stages of collaboration. Such MVPCs help get started with co-creation and create a sense of concreteness in a joint activity. MVPCs enable feedback provision from industry to academia in the early stage of the solution development, which increases the probability of creating the outcomes of mutual benefit. A minimal viable tool could be an artifact that allows automation of tasks that are normally performed by humans, or older and less optimal software tools. We emphasize the co-creation of tools as a way to encapsulate a solution for a given problem context, as opposed to cooperative method development (CMD) by Dittrich, in order to emphasize the creation of an asset that can be configured and executed. CMD on the other hand is less tangible and not an asset that is a manifestation of ideas into a software or physical object such as a sensor system. In C1, tools like TITAN and Depict were developed over continual interaction and testing with the respective industry partners. This lean approach to development gave a sense of ownership to the industry practitioners over the tools developed, and the tools adapted to suit their work environment. However, it is also very important to make sure the lean approach of development extends to more than just one partner. For instance, after being initially developed with the Norwegian Toll Customs, Depict was also improved due to tests with the database systems at the Cancer Registry of Norway. In C2, the Flow sensor developed by SZ provided only raw data from forces measured from the expansion and contraction of the ribcage. This was a minimal signal that could be used to create many new value propositions with many partners. This included prediction of sleep apnea, respiratory flow, and breathing patterns for stress reduction, to name a few. It would have been much more expensive and uncertain to launch a product in the market if SZ had to secretively develop it for a specific target over several months. In C3, researchers collaborate with practitioners to develop a MVPC on a publicly available dataset to predict tool wear in manufacturing. The publicly available dataset protects the industrial practitioner from sharing intellectual property while testing out ideas along with a researcher. The MVPC is developed to be generalizable to new datasets that will eventually be available from partners in the manufacturing industry. One such example of a co-created MVPC is the erroneous data repair system available on Github \footnote{https://github.com/ejhusom/Erdre}, which was demonstrated on the public data set on CNC milling tool wear \footnote{https://www.kaggle.com/shasun/tool-wear-detection-in-cnc-mill}.

\textit{Benefits:} Research value co-creation from the beginning of the collaboration, where industry gets to understand practical benefits of abstract research concepts developed at first by researchers, as well as to influence solution development by providing timely feedback. 

\textit{Related solutions:} P7, AP2, AP3.

\subsection{P7: Frequent Iteration}
\textit{Problem:} Progress on technical developments are reported sparsely, on each side of the collaboration. Consequently, useful feedback to guide further development is captured sparsely.

\textit{Context:} In technology projects where software tools are key artifacts in a collaboration, it is important to demonstrate progress frequently, and receive feedback to direct technology improvement efforts. Lacking feedback from researchers to practitioners makes practitioners disengaged. Lacking feedback from practitioners to researchers makes researchers base their solution on possibly inaccurate working assumptions (about the requirements and constraints for the solution needed by industry). Inaccurate working assumptions in software tool development can lead to technical choices which may not be easily correctable later on. For example, we encountered such an issue while developing TITAN in collaboration with Cisco Systems, where inaccurate working assumptions led to a limited choice of how the tool can be used. TITAN was designed as a standalone tool, while Cisco's preference was to use it as a service.

\textit{Solution:} Demonstrate progress and intermediate results of technical developments often. Researchers can obtain feedback from practitioners on the soundness of technology choices. Practitioners can keep researchers up to date on their technical developments, which is necessary for smooth integration of all technical artifacts developed. Frequency should be relatively high, daily or weekly, until a tipping point is reached, when the stakeholders using a co-created knowledge become self-sufficient. Ideally, the shorter this time is the better.  
The mutually acceptable frequency of iteration often reflects the interest level and investment of stakeholders in a project. It is necessary to take it as a cue and adapt accordingly. For instance, in C1, SRL was wary about introducing too many meetings that do not appeal to stakeholders or sometimes feel like a waste of time. Instead, the frequency was adjusted according to the level of intensity experienced in the collaboration. We recollect that the frequency increased when it was about trying tools hands-on and obtaining results, while it decreased when it was about writing scientific papers. In C2, the frequency was very high before an event where the sensors were demonstrated, such as the MusicLab organized by the University of Oslo or the World Rowing Masters regatta. The exhibition of results to a larger audience or public often creates more intensity of collaboration and frequent improvements. In C3, there is a high frequency of iteration in requirements specification and task allocation processes, due to the diversity in the partners. The partners are experienced in their respective domains and have tools available to treat manufacturing sensor data. Hence, frequent iterations entailed defining requirements and tasks for each partner, while minimizing conflicts of interest and ensuring the protection of intellectual property for commercial businesses. As a researcher at SINTEF pointed out \textit{"I think just because we are meeting so frequently, we avoid any conflicts over who is working on what".}

\textit{Benefits:} Bidirectional feedback provided often helps validate results often and ensure that potential misinterpretations and missed requirements are identified and fixed when the time is appropriate. This can be early on or when the tool needs to be demonstrated to a larger audience. Outcome technology developed becomes applicable, usable and scalable to the relevant context.

\textit{Related solutions:} P4, P6.

\subsection{P8: Two-faceted Problem and Impact}
\textit{Problem:} Problems addressed in a collaboration are not relevant for one side of the collaboration.

\textit{Context:}
Academic researchers and industry practitioners have different motives for collaborative projects. Researchers are driven by scientific challenges and practitioners by practical solutions. Although their drives are different in nature, for the collaboration to succeeded, all stakeholders need to have the opportunity to reach their objectives and create impact that matters to them.

\textit{Solution:} Define the problems with two sides to them i.e. as a practical problem and deriving from it, a research problem, which are closely linked. A practical problem is a need to improve unsatisfactory aspects of industry practice, caused by a concrete condition, for example, the lack of test automation. The practical problem caused by this condition may be the high cost of testing. A research problem is a gap in knowledge whose understanding is crucial for solving a practical problem. Thus by solving a research problem we create ground for solving its related practical problem. In C1, SRL defined the problem of test case selection and prioritization based on the needs of its industry partners ABB Robotics and Cisco partners. For researchers, this meant researching test optimization techniques, and for practitioners, this meant developing a tool that can alleviate the problem of manual test selection. As manual test selection was tedious, some of these tests were redundant and there was a need to research the problem of automatically selecting a minimal set of test cases and prioritizing them based on what changes were made in the code base. A researcher in C1 indicated that \textit{"A two-faceted problem definition allowed us to, on the one hand, advance the state of test optimization for highly-configurable software, while on the other develop a tool TITAN} \cite{b52} \textit{that reduced the effort of manual testing for Cisco Systems Norway"}. Similarly, in C2, researchers at the University of Oslo developed algorithms to predict sleep apnea from the breathing data obtained by the Flow sensor. This helped SZ define problems such as being energy efficient about overnight recordings, addressing  issues with Bluetooth connectivity and reliable data storage. In C3, the problem initially defined was to perform erroneous data repair. IDEKO did not see a benefit in solving this problem, although it was written as part of the description of work (DoW). Therefore, the problem specifications evolved slightly beyond the initial DoW. For instance, when SINTEF presented some of their previous work on data profiling with constraints on data, IDEKO became interested in using that technology in their manufacturing setup, as they saw it being immediately useful compared to erroneous data repair. SINTEF, however, initially followed the DoW, but soon realized that they can bring more value to the project through data profiling with constraints. Therefore, it was
necessary for SINTEF to be flexible and address the real industrial problem with higher priority that addresses a challenging research problem that was not immediately useful. Once trust is established through solving simpler and more pressing problems, it becomes easier to address more scientifically stimulating problems later on.

\textit{Benefits:} Both sides of IA collaboration see the benefits of solving joint problems, and consequently creating impact, each from their own perspective, i.e. academic and industrial impact. This condition greatly contributes to active commitment to the collaboration from all partners. 

\textit{Related solutions:} P1, P3, P11.

\subsection{P9: Joint Authorship of Scientific Articles}
\textit{Problem:} Industry practitioners are often reluctant to participate in joint authorship of scientific articles.

\textit{Context:} Authoring articles based on scientific methods combined with insights from industry experts can bring increased credibility and reusable knowledge from the co-creation process. However, an important challenge is to involve industry professionals in the cumbersome writing process that requires several rounds of proofreading, editing, precision in data and statistics followed by peer-review and publication. Industry practitioners are highly sensitive to the risk of revealing trade secrets and flaws in industrial systems through publications that could bring the business into bad light. Some have a strict internal publication vetting process before any form of external communication of scientific results. This often can discourage all stakeholders in performing thorough scientific studies. Nevertheless, transparency through published scientific articles about technology or methodology, such as high quality software testing practices, can bring trust in a company.   

\textit{Solution:} Writing and peer-review helps clarify complex concepts that companies often grapple with. Joint authorship of scientific articles must be seen as a way to distill complexity into a sequence of words and illustrations that clearly communicate the outcome of a co-creation process. Industry practitioners should ideally be engaged in describing their viewpoints in ways they are initially comfortable with. This can be in the form of user stories, answering a multiple-choice survey where the questions are carefully crafted, or interviews of experiences such as with software tools developed by scientists. The peer-reviewed publication and its acceptance by the larger community of researchers and practitioners can develop increased enthusiasm of industry practitioners in the scientific publication process. Increased motivation can lead to practitioners taking up industrial PhD positions in collaboration with a research group. For instance, in C1, an employee from ABB Robotics decided to realize an industrial PhD while on partial leave. During the PhD he published  scientific articles \cite{b47} in conferences and journals in the software engineering and testing community. Another practitioner from C1 pointed out that \textit{"It was only after we wrote the first paper together with researchers that we started appreciating such a writing exercise, as it helped us understand scientific writing, which is often a great source of innovative ideas"}. In C2, there were no instances of co-authorship, however, the industrial partner SZ has been acknowledged in several publications made by research groups, which gave credit to SZ and hopefully increased the company visibility. For instance, a study at the University of Oslo helped validate the accuracy of the Flow sensor developed by SZ in comparison to other low-cost sensors for sleep apnea studies\cite{b48}. In C3, some of the commercial businesses have an interest in publishing results from the project. For instance, PREDICT approached SINTEF to work on temporal clustering of manufacturing data with a publication in sight. The publication can document approaches implemented by PREDICT, who has limited resources for communicating their research to the public. Often ideas are lost in source code among companies that do not have the capacity to document and test their ideas. This, unfortunately, increases their technical debt in the project in the long-term. 

\textit{Benefits:} The consequences of joint authorship can have many advantages such as: a) documented industrial impact of scientific research, b) skill upgrade for industry practitioner, such as through industrial PhD programs, c) higher credibility for industrial products and processes due to transparency and evaluation in published scientific articles, d) more efficient use of resources because researchers can help practitioners write papers of high rigor, while practitioners can help strengthen the experimental evaluation by providing data and case studies, and e) increased ability of industry practitioners to absorb academic knowledge disseminated in the community typically through scientific papers.

\textit{Related solutions:} P5, P10.

\subsection{P10: Managed Intellectual Property Rights}
\textit{Problem:} The rights to the intellectual property created during a research knowledge co-creation need to be managed professionally to increase trust between partners.

\textit{Context:} Intellectual property is a category of property that includes tangible creations of human intellect. They can be categorized in the form of patents, copyrights, and trade secrets. For instance, in C2, SZ patented a sensor for respiratory inductance plethysmography. Similarly, partners in C1, such as Cisco Systems have several trade secrets in the domain of video conferencing systems. Academic partners, such as SRL, developed software programs that are automatically protected under a copyright as a defense mechanism against copyright infringement. Every stakeholder in a collaboration has an interest in intellectual property and the outputs of a co-creation process can generate property that some partners can exploit, as per a clearly specified agreement. However, as the observer in C1 noted down her observation from the observation sessions \textit{"Managing intellectual property is difficult, as it is very hard to trace where ideas originate from and how they evolve into tangible products."}

\textit{Solution:} A collaboration needs to be established based on a \emph{consortium agreement}, about who the producers and consumers of intellectual property will be, as well as how the intellectual property will be maintained, if required in the future. For instance, in C1, academic partners such as SRL are financed by the Norwegian Research Council through agreements. The consortium agreement stated that some partners, such as SRL, will publish research articles about their work with the industry after a careful scrutiny of the results together with the industry partners. Esito, for instance, was identified as an exploitation partner that would commercialize software, handle licensing agreements with other interested industrial partners, and provide maintenance support, if needed in the future. In C2, SZ obtained an US patent on their invention of the breathing sensor Flow. The patent allowed SZ to not only attract investors, but also have the possibility to license its technology to other companies.  
In C3, a consortium agreement has been signed by all partners on how intellectual property will be shared. All research results and ideas that are performed on the data made available by industry partners and that are accepted for publication in scientific journals can be used by all partners. However, there are unwritten restrictions and a mutual understanding among competing partner companies to not reveal too much to each other. Neutral partners, such as research institutes, play an important role in brokering the interaction and protecting the interests of competing industrial partners in a large project.

\textit{Benefits:} Establishing a legal framework such as a consortium agreement or a patent can simplify the collaboration and stem doubts that may arise around ownership of intellectual property.

\textit{Related solutions:} P9

\subsection{P11: Deriving Research and Business Critical Questions Come First and Data Sharing Later}
\textit{Problem:} Business critical questions for partners in a co-creation process need to be vetted out before seeking to curate data to address the questions.

\textit{Context:} Often researchers are eager to work on available data and generate insights.  
However, very often industrial partners need a good reason to share data, which is considered to be intellectual property and often can take several years to curate. Similarly, researchers themselves need to ask questions about which business questions are of scientific value and worth addressing before seeking to use whatever data is available to generate research questions and answer them. This also creates a bias to specify obvious questions that can be addressed based on available data. 

\textit{Solution:} Deriving research and business critical questions should ideally be performed before extraction of data. Here deriving questions entails discussions to understand which variables and metadata from a stakeholder can be of relevance to address a problem. This is in contrast to requesting all possible data and then formulating questions based on that. Data is usually an asset in a company and the sharing of data needs not be useful, unless the intention is clear. This also relates to the data minimization principle of the General Data Protection Regulation (GDPR). Data minimization is a principle that states that data collected and processed should not be held or further used unless this is essential for reasons that were clearly stated in advance to support data privacy. In GDPR, this is defined as data that is \textit{adequate}. To derive research and business critical questions, workshops where research and business questions are formulated, re-formulated and prioritized by stakeholders can help. However, it is assumed that stakeholders know what data is available and can help guide discussions towards formulating research and business questions that are within a given scope. What is the most important is the satisfaction of stakeholders in the formulation of the questions both for its business and scientific value. In C1, SRL organized workshops between stakeholders to come up with research questions relevant to businesses. These workshops intended to specify questions for research-based innovation. The questions need to be arguable with previous scientific evidence, if possible have quantifiable answers, and can be answered using prototypical development and experimentation. The questions should be leading a narrative, but also be mutually independent of one another. Once the questions were agreed on during these workshops, the rest of the half-year was used to co-create software prototypes to answer the questions. For instance, Toll customs asked the question: Can we verify that all customs declarations are sent to the statistical bureau of Norway before being archived? This led to the development of the tool Depict that was used to model and verify this requirement. In C2, SZ and UiO were interested in detecting obstructive sleep apnea from abdominal breathing data. Before the data collection took place it was very important to define research questions that needed to be answered. One of the first questions was about the quality of the breathing data and how good it is for clinical decision making addressed in \cite{b48}. Similarly, in C3, a research question of interest is what is the data quality of sensor data from machine tools. In particular, the completeness and consistency of the data. Addressing this question entails the development of data acquisition systems that ensure at least 99\% data completeness and consistency. At the beginning of the project, there has been a phase of gaining trust through presentations and refining research questions that can bring new value to industrial partners. For instance, an engineer from IDEKO explained \textit{"We use real-time systems to acquire high frequency vibration data from machine tools, but do not perform data quality checks. This is where we use the help of SINTEF to specify assertions on the data and to verify them in near real-time on data arriving at a frequency of 1Hz"}. This need had to be identified before SINTEF could gain access to their REST API to access data.

\textit{Benefits:} The benefit of deriving research and business critical questions first is to agree on topics and the use of time on something that is truly beneficial to stakeholders in co-creation. Also, it places much less pressure on stakeholders with data to give away the data without seeing the immediate benefits from it.

\textit{Related solutions:} P8.

\subsection{P12: Minimize moving parts}
\textit{Problem:} A co-creation project can have many moving parts such as a team size, partnerships, technology changes, and evolution of use cases leading to many source of uncertainty.

\textit{Context:} Both short and long term co-creation projects can have partners that suffer from employee churn leading to changes in team size, partners joining and leaving consortia, and evolution of use-cases with time. These moving parts can lead to uncertainty in terms of human resources and objectives of the project. If one partner is more prone to frequent changes this can also amount to  deterioration of trust with other more stable partners. 

\textit{Solution:} Minimizing moving parts in a project is typically necessary for its stability and longevity. When dialogues are established between specific people it is best to select the compatible people and maintain the dialogue instead of replacing people. Changing personnel in a dialogue can entail a large cost in on-boarding and re-establishing trust. If a company is large enough it may even be necessary to have at least 2 employees following up on a project (bus factor of 2), such that the project stays on track even if one of them leaves the company. Similarly, in terms of use cases and their owners it is ideal if partners and problem definitions stay stable during the course of the project with minor variations. Choice of technological solutions to implement in a project should have a good mix of stable and experimental elements. The experimental aspects with high activity could be carried out in parallel, without affecting the long term goals of a project hinged on more stable technological solutions. In C1, SRL relied on developing tools based on the stable Eclipse framework. The experimental elements were built as extensible plugins in the project without affecting the base software. Both TITAN \cite{b52} and Depict \cite{b44,7302459} were tools developed on the Eclipse platform. Most employees recruited on the team were PhD students or postdoctoral fellows with a work contract of 2-3 years enabling long terms stability in the project. In C2, a handful of critical members were hired full-time by SZ to ensure sustained development and customer support for a resource constrained setup in a startup. These included two software developers, one industrial designer, one electronics engineer, and the CEO. The other employees were only hired part time and some of them served roles in other organizations. In C3, efforts from partners are bounded by the number of person-months. There are usually 2 or 3 members from each partner attending all meetings in the project. The meetings minutes are documented and many of the meetings are recorded after the start of COVID-19. The dialogue and relationships between partners have matured due to conversations among the same people for a long period of time.

\textit{Benefits:} As a project manager in C3 summarized, \textit{"Minimizing moving parts brings stability to a collaboration and allows going deeper into solving a problem by a team. It also helps minimize the unpredictable costs from on-boarding new members, learning new and experimental software tools, and dealing with changes in use cases under study."}

\textit{Related solutions:} None.

\subsection{P13: Anticipate Unavailability of Data}
\textit{Problem:} Researchers are eager to obtain data and perform statistical analysis, machine learning and generate visualizations to bring valuable insight. However, very often the data obtained from partners may not be conducive to statistically significant conclusions or adequate for tasks such as machine learning.

\textit{Context:} When data is collected from an observation it is not always stored with a clear purpose. Some variables can be observed and their evolution over time is stored for potential future use. For instance, data can be in the form of logs while executing a software program. These logs represent daily use, but are not necessary obtained from a controlled experiment. Using such data for making scientific conclusions may not be easy, as it does not allow clear comparison between two different scenarios, such as in a randomized controlled trial. Many companies may collect years of such data and expect some value from it. Similarly, researchers seek high quality data amenable to making scientific conclusions with strong statistical significance. The mismatch in data made available and its expected utility for generating scientific evidence is a problem all parties in a co-creation should be wary about.

\textit{Solution:} It is important to anticipate unavailable data necessary to generate scientific evidence and plan for it. This can be by: (1) defining scientific problems that can be addressed satisfactorily with existing stream of data, or (2) specifying and carrying out a new data collection protocol under controlled conditions amenable to high statistical significance. In C1, researchers decided to define scientific problems based on available data. For instance, in a collaboration with Cisco, researchers were provided with daily logs from automated testing. In \cite{Sharif2021DeepOrderDL}, SRL researchers use deep learning to help find the most failure-prone test cases in continuous engineering processes at Cisco Systems, without enforcing a new data collection protocol. On the other hand, studying the impact of the mobile game intervention FightHPV\cite{ruiz2019fighthpv}, developed in collaboration with the Cancer Registry of Norway, required a stricter data collection protocol, where data about people using the app was collected based on informed consent, using their electronic national ID, also called BankID. Data about how people used the app and the learning from it were used as part of an observational study comparing the uptake of screening and HPV vaccination before and after the launch of FightHPV.   
In C2, SZ required data from the Norwegian Institute of Sports Science to create physical and deep learning models to predict respiratory minute ventilation from ribcage movement data. Since the data collection process is cumbersome, the development of the deep learning models was based on maximizing the use of deep learning techniques that can generalize reasonably well from small data and discuss the risks of faulty prediction for unforeseen data \cite{DBLP:conf/ijcai/SenBH20}. In C3, sensor data from machine tools such as CNC milling, turning, broaching, and grinding is needed to develop tools to validate and improve data quality for Industry 4.0. The sensor data from machine tools is sometimes not available due to technical issues such as variable frequencies, and sometimes due to security issues preventing data tampering. Our strategy to address this lack of data is to summarize data based on top five frequencies and discrete machine states derived from a high frequency acquisition of accelerometer data at the edge that is unavailable. This summary data is secure and contains adequate information to derive data quality hallmarks by other partners in the team.

\textit{Benefits:} As a researcher in C3 mentioned \textit{"Anticipating the unavailability of data is very important to manage expectations in a team"}. It is also important for thinking creatively about generating value from available data or establishing a data collection protocol very early in the process, in order to obtain statistically significant results.

\textit{Related solutions:} None.

\subsection{P14: Define long-term tasks that enable remote collaboration}
\textit{Problem:} Continuous engineering practice used in the industry with short cycles to release is also often expected from research activities.

\textit{Context:} Continuous engineering practices in the industry are based on rapid releases of product updates with a customer base and utility that is relatively well-defined compared to a co-creation process involving research. This agile approach is beneficial when goals are clear and tasks are well-defined. However, problems that require long-term research and experimentation are often expected to be broken down into small time-bounded tasks, as seen in continuous engineering best practices. This is, however, not always feasible for every problem where ideas need to mature from uncertain and unpredictable experiments carried out by collaboration between researchers and industry practitioners.

\textit{Solution:} Researchers need to study existing practices and read  recent research and think in isolation to have mature revelations that can help industrial stakeholders think out of the box. Research activities need to be formulated as long-term tasks running in parallel and should be enabled by remote collaboration to draw on minds and expertise from around the world. Remote collaboration is even more pronounced during the COVID-19 pandemic which has forced many teams to collaborate remotely and has become the norm. This is especially necessary for members of the collaboration obtaining a PhD degree or undergoing postdoctoral training. On the other hand, industrial partners focus on short-term tasks that have a granularity small enough to be completed in a two-week sprint. This makes it difficult for industry to think long-term and hence it is a perfect marriage and a win-win situation to work with neutral researchers who can look into topics that require long-term thinking and experimentation. In C1, SRL researchers explored long-term research topics such as the use of deep learning for test case prioritization and selection in continuous integration testing at Cisco Systems \cite{Sharif2021DeepOrderDL}. The researcher was a PhD student whose work helped optimize nightly build and test runs of Cisco's conferencing systems. In C2, SZ had the long term goal of developing a mathematical model to predict respiratory flow from ribcage movement data obtained from their Flow sensor. This work required collaboration with the Norwegian School of Sports Science to obtain data to verify the validity of a physical model and also test new techniques in deep learning \cite{DBLP:conf/ijcai/SenBH20}. Data acquisition from an exercise spirometer and the Flow sensor developed by SZ was a long-term task that was performed in parallel and remotely by the sports school. Meanwhile, SZ was focused on selling their sensor devices to stakeholders who could build their own applications. In C3, most manufacturing companies have a continual production of parts and have been engaged with researchers to use the data from sensors monitoring the manufacturing process. For instance, RENAULT produces 1500 cylinder heads for their Cleon engine per day. Their CNC machines can generate several hours of high frequency sensor data while the production is ongoing. Researchers from SINTEF use this stream of multivariate sensor data to do long-term research, i.e. compute metrics for data quality to ensure that it has a completeness and consistency close to hundred percent. The data quality research is running in parallel, without interrupting the manufacturing process. The data quality research is in turn used to attain the long-term goal of zero defect manufacturing. As a researcher at SINTEF said \textit{"Long-term tasks facilitate remote work, as we work remotely in Norway and collaborate with partners in France and Spain to acquire relevant data for long-term research."}

\textit{Benefits:} Defining long-term tasks allows stakeholders to perform activities that require maturity. It also aligns well with the goals of young researchers who wish to establish themselves over a period of 2-3 years (PhDs and postdocs), not just as problem solvers but also thinkers and philosophers in their domain. Long-term tasks also facilitate remote work that is necessary for isolated thinking and reflection, and has become inevitable during the COVID-19 pandemic.

\textit{Related solutions:} P4, P7.

\section{Anti-patterns for Industry-academia Collaboration}
Anti-patterns are an extension to patterns. Anti-patterns are frequently used practices to commonly occurring problems that are ineffective \cite{b36}. They are useful because they describe the causes of failures, which are important to understand, so that the corresponding failures are not commonly repeated but are avoided and mitigated. Similar to patterns, anti-patterns are specified using templates, which supports their reusability. We use the following template: \textit{\textbf{Synopsis \ding{169} Context \ding{169} AntiPattern Solution \ding{169} Symptoms and Consequences \ding{169} Refactored Solution \ding{169} Benefits and Consequences \ding{169} Related Solutions}}, illustrated in Figure \ref{fig_template}.
Anti-patterns are related to a specific \textit{problematic solution}, which occurs in a specific \textit{context}. They could be patterns which have over time become problematic, or which are not working in a context other than initially proposed for. It is important to document how the problematic solution became problematic by incorrectly resolving underlying problems that exist in the context. \textit{Symptoms} are useful for recognizing a specific problematic solution (anti-pattern), and understanding its implications. \textit{Consequences} describe the implications of the problematic solution. \textit{Refactored solution} is an effective method of resolving the problematic solution. Refactored solution provides \textit{benefits}, but it may also have remaining unresolved issues called \textit{consequences}. Finally, the refactored solution can be related to other \textit{related solutions}.

\subsection{AP1: Miss the forest for the trees}
\textit{Synopsis}: Focus on individual evaluation parameters and expect the emergence of impact.

\textit{Context}: The success of IA collaboration project is typically measured on each side of a collaboration using specific metrics. Such measurements help organizations evaluate the benefits of investing resources into such types of projects and serve as compass for future such investments.

\textit{AntiPattern Solution}: During collaboration, scientists are asked to produce output such as high number of publications, successful grants, and a number of graduated PhD students. The term h-index is often used to measure impact as the maximum h number of publications cited h times. However, there are many ways to boost h-index, which include self-citation, or obliging authors to cite papers during a blind review process, or collaborating with mature authors, to name a few. Similarly, industry practitioners are interested in sales and profit, as their bonuses are linked to such numbers. In addition, they overlook the ethical and environmental impact of their products and services.

\textit{Symptoms and Consequences}: Focusing on parameters that evaluate scientists and industry practitioners individually often takes the focus away from the real expected impact that can be achieved as a team. Francis Darwin, the son of Charles Darwin once said \textit{"In science, the credit goes to the man who convinces the world, not to whom the idea first occurs"}. This quote succinctly summarizes the difference between impact and output. 
Individual evaluation parameters may be well perceived for an individual's career progression, but it is not necessarily a good indicator of impact in our society. Boosting individual parameters in academic excellence may overshadow innovative ideas among younger researchers. 
Many potentially world-changing inventions taking birth in academia end up in the so-called technological valley of death as researchers lack the skills or support to convince funding agencies that their inventions are worth investing in and can change the world. Similarly, many companies die both slow and quick deaths by focusing only on short term goals, such as sales, while ignoring ideas developed in research labs and startups or societal trends.

\textit{Refactored Solution}: High impact is defined by the Cambridge English dictionary as the ability to
withstand great force. Although this refers to materials, the same concept can be applied to scientific research or industrial products. Research that can stand the query and test of the brightest mind outside the echo chambers of confined research communities has the potential for impact. Similarly, products and services offered by companies need to be able to beat their competition in several aspects to achieve real impact. High impact can be achieved when academia and industry can pinpoint the improvement or creation that end users see as a welcome change. For example, in C1, one of the industrial partners, Cisco Systems, was interested in building the highest quality of video conferencing systems with special emphasis on security for high-end conversations. This drove them to make their testing methods of world class standard, in collaboration with SRL researchers. A number of articles published on testing video conferencing systems increased the trust in the quality of Cisco's products. In another C1 project, a project manager at Toll Customs said \textit{"We call this a high impact project; we see the improvement in our daily practice by using Depict, and we have joint publications showing this".} In C2, the impact at the startup phase of SZ translated to increased sales in sensors. SZ achieved impact by collaborating with numerous partners and having them purchase sensors and use them in their research. The published results and communication led to increased contacts from more individuals, companies and universities worldwide for sales of sensors. SZ collaborated with artists and studios to promote the technology through installations in galleries and the public space.  
These activities generated media coverage and direct interaction of SZ's technology and the general public. 
In C3, it is important for individual partners to have well-defined problems to solve.  
In monthly meetings, we constantly re-align the partners by asking them how their work contributes to the KPIs of 99\% average completeness and consistency in manufacturing data quality. These metrics can be realized individually, however, continuous re-alignment of the partners' metrics increases the quality culture in the entire manufacturing line. As operators are made aware of the quality of the data that is acquired, it gives them higher confidence in perceiving quality. This helps enable the ultimate goal of zero defect manufacturing and reduction of industrial waste addressing UN's sustainability goal 12 on responsible consumption and production.

\textit{Benefits}: Greater impact creation as a result of focusing on the bigger picture of how co-creation in a team is bringing real change in the society, rather than on individual evaluation metrics. 

\textit{Related Solutions:} P3, AP13.

\subsection{AP2: Non-reusable minimal viable solutions} 

\textit{Synopsis:} Developing solutions with minimal features for problem owners to validate the feasibility of ideas quickly, but overlooking solution extensibility and sustainability.

\textit{Context:} It has become a common mantra to develop minimal viable software products in projects to impress and engage stakeholders. 

\textit{Symptoms and Consequences:} Many IA collaboration projects need to be realized in short periods of time, resulting in the development of minimal viable products. These products often end up becoming project-specific. There may be similarities between such projects that run simultaneously, but since they last for a limited time, there is very little time to abstract from the different repositories and develop coherent (and possible more generalized) artifact that could be reused. Therefore, there is a risk of reinventing the wheel for each project in the worst case scenario. 

\textit{Anti-pattern solution:} A minimal viable product or MVP is quickly developed by an academic group to address an immediate industrial need. It is often done with impressive speed by reusing open source solutions and existing long standing libraries and tools. The development of an MVP is a well-known concept in the lean philosophy for startups and has found its way into academia.

\textit{Refactored Solution:} Research groups need to spend time in developing and maintaining a vision and software artifacts that can stand the test of time. It is important to develop a core \emph{engine} that can interface to many sources and sinks, and can be deployed on many operating platforms, with features that represent building blocks for constructing solutions for projects with different requirements and stakeholders. The abstraction of what can be relevant for a long time requires continuous dialog and exchange of ideas between different project that execute simultaneously. The goal is to enable abstracting from artifacts such as prototypical source code developed in projects. However, such communication between projects, especially spread in time, is also contingent on the clauses of an intellectual property rights agreement (if one exists). It is also necessary to build software artifacts with an architecture that is easy to use and extend. In C1, a core engine for test case optimization was developed that was applied to problems presented by several industry partners, including Cisco systems and ABB Robotics. In C2, SZ developed a skeleton app with an extensible architecture that could be easily adapted and tested for the needs of various stakeholders in a short period of time. A software developer at SZ indicated \textit{"Developing an extensible app saved us so much time when reusing the solution for different applications"}. In C3, all software artifacts are developed with extensibility in mind. For instance, SINTEF developed a system for erroneous data repair that is based on a data pipeline that can be improved based on new machine learning models or new components, developed by other project partners, such as for explainable AI. 

\textit{Benefits:} Moving from the idea of a minimal viable product to the concept of developing a \emph{minimal extensible engine} can greatly improve reuse and applicability to a number of diverse problem domains. This mindset helps a research group position itself as the maintainer of a constantly reused software artifact, giving the group recognition in its community.

\textit{Related solutions:} P6, AP3, AP8.

\subsection{AP3: Premature results in short-term collaborations}

\textit{Synopsis:} Short-term IA collaborative projects may pose lower risk for industry investments, but inevitably lead to limited practical and scientific impact.

\textit{Context:} Industry side of IA collaboration typically prefers short-term projects compared to long-term ones \cite{b42,b43}, as a way to, for example, cater to requirements for a short-horizon investment of resources. 

\textit{Anti-pattern Solution:} The short term may be aligned well with the goals of most industry companies, for instance, it speeds up learning about whether investing in the collaboration is worthwhile. However, short-termism is not always in alignment with a long term research vision of the project. 

\textit{Symptoms and Consequences:} In such short-term collaborations research goals are often left yearning for scientific abstraction and maturity. A researcher at C1 indicated \textit{"Our students often struggle with publishing the results coming from short-term projects, because they have not reached the maturity required for publication"}.

\textit{Refactored Solution:} Projects aiming to achieve both practical and research impact should aim for a mid-term duration. This will give enough time for the thinking and abstraction process that is necessary in scientific research, but will also allow the creation of time-tested tools that industry can effectively apply and assimilate. From our experience, both C1 and C2 collaboration contexts benefited from structuring longer projects in sprints defined over a period of maximum three months, before they are re-evaluated and adjusted. This was an optimal way to keep the project partners engaged in short bursts of a few months. Scientific maturity was attained through incremental publication and going through a peer-review process every six months or so. The scientific publication process gave us external validity for our chosen direction, while also helping us address questions that were not part of our foresight. This led to more maturity in the scientific aspects of the collaborations. In C3, certain deliverables have a very short time to be developed and accepted in the early stages of the project. It is often important to communicate the premature nature of some of these deliverables to reviewers. Moreover, the project is planned in an incremental manner, such that new versions of deliverables are produced every year and the quality and maturity of the results from the project improves with time. 

\textit{Benefits:} Achievement of satisfactory levels of both scientific maturity and practical readiness of the results created in a collaboration.

\textit{Related Solutions:} P6, AP2.

\subsection{AP4: No skin in the game}

\textit{Synopsis:} "Free" IA collaboration projects may attract industry partners, but are amenable to the risk of insufficient investment from industry partners and overall dissatisfaction. 

\textit{Context:} Funding programs for IA collaboration projects involve academia and industry at varying levels of investment from each partner.

\textit{AntiPattern Solution:} In some IA collaboration projects, no investment is required from industry, thus making such projects "free" for industry. Researchers spend several months to identify a relevant problem and come up with a solution that can benefit the industry partner. However, the industry partner makes limited involvement in the project.

\textit{Symptoms and Consequences:} The consequences of a partner not having skin in the game is less intensity and involvement in a project and a constant feeling of fragility. The partner that has more skin in the game will have to go out of the way to make sure the partnership holds with partners that have less at stake. The success and longevity of a relationship is also in jeopardy when partners give less and get services for free.

\textit{Refactored Solution:} "Skin in the game" is to have incurred risk (monetary or otherwise) by being involved in achieving a goal. It is important for partners to have skin in the game in a co-creation project, such that they incur a cost for, for example, leaving the project midway. However, it is difficult to enforce such costs for long-term projects spanning several years because relationships, trends, and enthusiasm can greatly vary. Hence, such costs should only be associated with short-term projects of 3 to maximum 6 months. The simplest approach to achieve skin in the game is by introducing "in-cash" investment into a collaboration, so that all parties have a vested interest in the project. However, in-cash investment also has the potential of souring relationships between partners, as trust may be lost and one partner may want to pull the collaboration in a different direction. In C1, SRL established a consortium agreement where partners had to commit to at least one person-year every year in a project for an 8-year project. In C2, partners were required to purchase sensors developed by SZ in order to obtain support. The alternative of lending sensors did not always work very well, as some partners did not have a concrete plan on how to use them. A project manager at SZ noted \textit{"Once we decided to put a price on the sensor, we could clearly see who of our collaborators is truly interested in collaboration"}. In C3, most partners are invited into a consortium based on prior experience of working together. Partners who do not meet their commitments often undergo a natural selection process of being excluded from funding opportunities in subsequent projects. Hence, it is of paramount importance for both industry and research to attain a level of satisfaction in the collaboration by delivering what is promised. Skin in the game is ensured through the monitoring of efforts made in deliverables, attendance lists and the level of participation in meetings.

\textit{Benefits:} Having skin in the game will change the way partners collaborate, taking joint ownership of the project. It may also mean that there can be intellectual property disputes for the work put into the project.

\textit{Related Solutions:} P1, P5, P8, AP5.

\subsection{AP5: Inequitable value creation} 
\textit{Synopsis:} Optimizing key performance indicators to create value, and thus engendering equitable value creation.

\textit{Context:} Researchers and industry stakeholders in a project are measured based on key performance indicators that quantify the value they generate.

\textit{AntiPattern Solution:} Value creation for  partners is an essential aspect in a research based co-creation process. Value creation is any process that creates outputs that are more valuable than its inputs. This is the basis of efficiency and productivity and is often quantified as key performance indicators (KPIs). Value for businesses can be quantified in the form of KPIs such as number of improved processes, products, sales, and services. Value for researchers can be quantified in the form of KPIs, such  as publications in journals with high impact factor, creation of open-source tools and open data sets for research.  It can also be seen as enhancement of knowledge or well-being in the workforce for all partners. 

\textit{Symptoms and Consequences:} There can easily be a discord in which path creates value for each partner. For instance, researchers may be too focused and isolated while writing scientific publications, neglecting the need for value creation for an industrial partner, which may entail improving a product that is used on a daily basis. Similarly, industry partners may not see the value in a validation study of their products and services, which can create more trust in their company among shareholders and customers.

\textit{Refactored Solution:} Value creation must be discussed with clarity in the very start of a project. Value creation can take place both in the short-term of a few months and the long-term over a few years. Value creation must be quantified such that each partner comes out of the project feeling that their inputs resulted in a something of value to themselves. There must be an equitable balance in value creation for all partners. For instance, if researchers obtain a high impact publication, enough effort and time must be spent in ensuring that the result also increases sales or leads to improved products and services for industrial partners. Quantifying such creation of value over a common index is a possible way forward. In C1, SRL defined value creation in terms of a balance in invested effort in the project, along with metrics about artifacts created by different partners, which include reports, research papers, software, or new sale leads. These metrics, when seen together, gave the consortium an idea on where to invest more effort to keep the value creation equitable. In C2, equitable value creation was measured in terms of markets made available to SZ based on the research collaboration. For instance, prediction of obstructive sleep apnea by University of Oslo opened a very large market that helped SZ refine its business strategy. In C3, making value creation equitable entailed finding a balance between creating data quality tools for manufacturing partners and publication of scientific results. The scientific results were essential for the research institutes such as SINTEF and TU DARMSTADT, while the tools developed during the project would eventually become prototypical components in the product lines of partners in the manufacturing sector, such as IDEKO, PREDICT, and DANOBAT. 

\textit{Benefits:} Equitable value creation in a co-creation project can increase its sustainability and strengthen the cohesion in generated value. For instance, establishing a causal link between high-impact research resulting in increased visibility and sales, and improved stock valuation for an industry stakeholder can be a mutually acceptable win-win situation. A researcher at C1 said \textit{"Equitable value creation is a must for sustainable collaboration, otherwise, one side of the collaboration will give up"}. 

\textit{Related Solutions:} AP4.

\subsection{AP6: The Incomplete Tower of Babel}

\textit{Synopsis:} Industry and academia speak different languages, especially differing on degree of mathematical formalism, which leads to less clarity in co-creation.

\textit{Context:} Stakeholders do not speak the same language, like in the biblical story of the Tower of Babel, which stayed uncompleted because the workmen could not understand one another's language. Industry practitioners and academic researchers are groomed in different environments, from which they inherited different vocabulary, with different degrees of formality. However, when practitioners and academics are involved in a coordinated effort to solve a problem together, it becomes important to communicate ideas in the same terminology, in order to avoid confusion and ambiguity, and to make sure that arguments and idea are conveyed in a clear and convincing way. For instance, from our experience, the term testing for academics is less about physical and real-world testing (which is how industry usually interprets it) and more about generating variations of test cases in a constrained domain. Moreover, what are software tools for academics, are only prototypes for practitioners, which lack robustness and many important -ilities \cite{Voas}.

\textit{AntiPattern Solution:} Academics like to strip technology off the hype and specify a well-defined problem in the language of mathematics, while industry practitioners build their careers on technologies that are contemporary and cool. As a software engineering practitioner in C2 pointed out \textit{"When we explain our daily practice to our academic partner, it is very difficult to use the same language they do, so they can understand us"}. In a collaborative project, academics talk to industry practitioners, make notes and create a mental image of the problem. After that, academics typically isolate themselves and transform the mental model to a model in a mathematical terms and formulas. They essentially convert an ill-conditioned problem from the real-world to a well-conditioned problem using their preferred mathematical formalism. This formal specification is then used to exchange data and perform experiments, to solve a problem that concerns the industry practitioner.

\textit{Symptoms and Consequences:} The differences of terminology and language between academics and industry practitioners distance the stakeholders over time. A lack of dynamic and business-focused communication between academia and industry often leads to scientific research straying too far away from the changing needs of industry practitioners.

\textit{Refactored Solution:} Understanding provides the foundation for effective collaboration. Therefore, both parties in collaboration must make an effort, from the early stages of collaboration, to understand the terminology used by the other party and be \emph{synchronized}. This can be done spontaneously in a day-to-day interaction, but also focused on business purpose, through lightning talks on specific topic. Such approach helps readjust academic language to continually evolving industry needs, while practitioners learn to think clearly in terms of well-defined mathematical formalisms. For example, in C1, SRL researchers made short talks on the topics relevant for the work being done, such as combinatorial testing or multi-objective optimization along with a presentation of how a real-world problem may be formalized. Practitioners, such as from ABB Robotics and CISCO Systems, demonstrated how they test product features, which was eventually formalized as a combinatorial interaction testing problem. The important aspect of such knowledge exchange is asking frequent questions and clarifications, instead of assuming what the term means. SRL and its partners spent a large amount of time understanding and defining what testing really meant for each partner. For instance, testing for the Norwegian Toll Customs entailed verification of databases consisting of customs declarations. While testing for ABB Robotics meant adherence of the operation of a painting robot to a trajectory specification. In C2, there were differences in terminology between machine learning experts and those who were performing signal processing on the sensor data. For instance, signal processing experts use techniques for transforming signals in a time domain to a frequency domain, to obtain different features for reasoning about sensor data, while machine learning experts directly use raw time series data and a neural network to obtain outputs of interest with minimal feature engineering. We bridged the gap between the two schools of thought by comparing results from both approaches, as part of the software pipeline from early stages of the project. In C3, partners in the manufacturing domain and those in ICT often speak different languages. For instance, SINTEF formulated the problem of erroneous data repair in manufacturing sensor data as a machine learning and software engineering problem, involving concepts such as data version control, data pipelines, convolutional neural networks, epochs, training and test sets, and feature engineering. Despite the mathematical formulation of the machine learning problem, it was made quite clear to the stakeholders in the manufacturing domain how the output of a neural network will contribute to improved data quality. 

\textit{Benefits:} Continuous synchronization between academia and industry to improve mutual understanding of concepts, models, and experimental results can help both academics and practitioners to achieve a higher level of clarity in co-creation.  

\textit{Related Solutions:} P4.

\subsection{AP7: Immeasurable objectives}

\textit{Synopsis:} Objectives serve the purpose of informing the partners of what can be expected from the collaboration. However, immeasurable objectives serve no purpose, other than to contribute to discontent of the stakeholders.  

\textit{Context:} Measuring the progress of a project at different stages of an IA collaboration is a good practice that helps reach the project goals faster. 

\textit{AntiPattern Solution:} When the problem analysis is conducted at a too high level and requirements of the target solution collected casually, it is difficult to define clear and measurable objectives. The example of an unclear objective is \textit{to develop an effective tool for test case prioritization}. The problem with this objective definition is that it is too vague and unspecific, providing no means to measure when the objective has been met. 

\textit{Symptoms and Consequences:} Ill-conceived objectives nearly always guarantee dissatisfaction with the outcome for the problem owner.

\textit{Refactored Solution:} It is necessary to analyze the problem to be solved thoroughly, to capture detailed requirements, and to understand the desired outcome. This information should then be translated to an objective in a formalism with well-defined syntax and semantics (e.g. logic, mathematical function). Objectives need to be specific and measurable. In C1, for instance, we defined the objective as \textit{to develop a tool that applies test case prioritization based on high fault detection as an objective function}. The effectiveness of the tool can be measured in terms of fault detection capability of prioritized test cases, compared to manual test selection, using historical test data. In C2, we specified an objective as a machine learning problem. \textit{Given overnight breathing patterns with manual annotation of apnea events, create a machine learning model to automatically predict sleep apnea from new overnight breathing patterns}. The effectiveness of the objective can be measured in the form of prediction accuracy and recall, and represented as a receiver operator characteristics (ROC) diagram. As a researcher in C2 observed \textit{"These measurements give clarity and confidence to a collaboration, as they serve as scientific evidence."} In C3, it was necessary to break down the high-level KPIs of achieving 99\% data completeness and consistency to smaller manageable and measurable goals. SINTEF organized presentations for each partner in the group and produced a synthesis of who does what with the manufacturing data. This information was then used to break down the work into time-bound goals for each partner, to improve data quality and help achieve the goal of 99\% completeness and consistency. For instance, SINTEF will develop a module for data profiling and erroneous data repair, INLECOM will develop a module for anomaly detection, PREDICT will develop a module for temporal clustering of manufacturing data, and DNV GL will develop the architecture for data quality as a service. All these individual efforts have a measurable impact on the KPIs.

\textit{Benefits:} Well-defined objectives lead to measurable progress and more satisfaction with the results produced.

\textit{Related Solutions:} P11.

\subsection{AP8: Confound lab setup with the real world}

\textit{Synopsis:} In software engineering, only the research knowledge and prototypes tested in the wild stand the chance of success when deployed to operation.  

\textit{Context:} To prove useful in practice, the results of IA collaborative projects need to be thoroughly tested for its envisioned application scenarios. 

\textit{AntiPattern Solution:} Software engineering researchers often work to solve practical problems. For example, in our collaboration with Cisco, we used test logs to improve the cost-effectiveness of testing video conferencing systems. In the collaboration with ABB, we used operational data to predict maintenance activities of ABB robots. When designing solutions to such practical problems, researchers may overlook the requirements and constraints of the solution to be deployed in practice, such as the volume of data being produced daily, or they may test the solution in overly simplistic scenarios. Failing to take real requirements into consideration results in a limited scalability of the candidate solution. What seemed a promising technology when tested in the lab suddenly cannot be transferred into a working solution in practice.

\textit{Consequences:} Knowledge developed and tested in the lab does not generalize to the real world environment.

\textit{Refactored Solution:} During early stages of collaboration, researchers need to understand as much of the domain as possible, to identify the technical and non-technical requirements of the solution sought. Because the requirements form the basis for deriving working assumptions that define what the target solution should look like, early requirements capture is crucial. In addition, later, during the solution testing, requirements can be used to generate a range of more realistic test scenarios to be checked, to validate the robustness of the systems once deployed in operation. In C1, SRL came close to recreate a real world setup, given budgetary restrictions. The research lab acquired a video conferencing system from CISCO Systems and an UR3 collaborative robot from Universal Robotics to perform testing research. This industrial equipment follows the same standards as higher-end versions in the product line of the partners. Performing experiments with equipment was transferable to a real world setup in a factory, due to the use of the same standards and software programs. In C2, sensors were tested at low-temperature ice baths, and for waterproofing we used a small pressure to ensure that its connectivity and thermal sensitivity is understood and managed before being deployed to a partner. These experiments were repeated several times to ensure that there were no confounding factors in the measurements. In C3, initial experiments on data quality validation were performed using a publicly available dataset on CNC milling \footnote{https://www.kaggle.com/shasun/tool-wear-detection-in-cnc-mill}. However, demonstrating the algorithms for data quality validation and repair on this dataset led to obtaining access to IDEKO's live CNC machine with real data from production operating 24/7.

\textit{Benefits:} Research knowledge and prototypes, when deployed in practice, scale to the complexity of the real world environment. A software engineering practitioner in C1 said \textit{"We seamlessly deployed the tool in our test framework, largely because the tool was developed and tested by working with real video conferencing system and data."}

\textit{Related Solutions:} AP2.

\subsection{AP9: Sunken cost fallacy in public research grants} 

\textit{Synopsis:} Investments in IA collaboration projects should not be contingent on previously invested resources, but on a clear sign of the collaboration sustainability and ability to create industrial and societal impact in longer term.

\textit{Context:} Academia aims to obtain long-term public grants (3-10 years) from national research funding agencies or the European Research Council for financial stability. 

\textit{AntiPattern Solution:} Academic researchers create consortia with industrial partners and apply for funding from national and European research councils. These grants are aimed to support long-term value creation and knowledge transfer from research labs to the industry. The research councils often tend to reinvest in the same people and companies and research groups over several years or sometimes decades.

\textit{Symptoms and Consequences:} The model of supporting businesses and academia with public grants has been applied for  decades  in several countries with reasonable success in terms of exploitable intellectual property and improving training of young professionals prepared for industry. However, many research programs continue to get funded because they have received funding previously. The funding is often obtained by the same person growing in a leadership role. This prevents new and innovative ideas from getting enough attention and funding to develop, because the priorities, vision and tone are set by the leader. Moreover, there is little accountability in terms of industrial and societal impact that is financially sustainable without public grant support.  

\textit{Refactored Solution:} Research groups should address industrial and societal impact and its financial sustainability during a long cycle of funding. This can involve taking risk such as spinning off a business with private investments to demonstrate adequate skin in the game from various stakeholders, innovation through patents, and job creation. This effort should be recognized by the public grants bodies and used as evidence for a new round of research funding. In C1, SRL had one attempt of spinning off a software testing startup based on the technology developed through the research collaborations. However, due to lack of resources, the initiative did not continue. A researcher in C1 who was leading the spin-off initiative reflected that \textit{"While eventually we did not manage to get the startup off the ground due to the lack of resources, through the venture with investors we proved the innovation potential underlying the technology, which was followed by some seed money"}. In C2, maintaining financial stability required SZ to pivot from a consumer product company to a consulting firm that allows white labelling of its sensor product to larger actors. SZ did not actively seek public grants due to large effort and relatively low chances of success. It is important to also note that the creation of business entities requires that people with business, marketing, sales and design acumen meet scientific researchers and that such a meeting place is provided. In C1, SRL established the Simula Garage, an incubator to facilitate such interactions. In C3, the ongoing project is a follow-up of another EU project MC-Suite \footnote{http://www.mc-suite.eu/} that was of smaller size than InterQ, in terms of budget and the number of partners. InterQ had to mitigate the perception of financing the same partners in MC-Suite and involve new partners that they have not worked with earlier. The addition of new partners, such as TU Darmstadt, INLECOM, and DNV GL has helped bring in many new ideas into the project that were not foreseen by the older members of the project. 

\textit{Benefits:} The benefits of nudging use-inspired research and co-creation towards financial sustainability will help minimize the dependence of research groups on purely public funding. Public grant bodies on the other hand can diversify their investments and avoid the sunken cost fallacy.

\textit{Related Solutions:} AP11.

\subsection{AP10: Adhering to traditional reward mechanisms} 

\textit{Synopsis:} Adherence to traditional reward mechanism may be suited to individual partners, but only adherence to bilateral reward mechanisms can bring about successful IA research co-creation. 

\textit{Context:} All partners in a co-creation project are driven by a traditional reward systems internalized from years of operation in specific contexts. These reward mechanisms are followed to achieve a lot of activity and quantification of numbers that help individuals advance in their careers.

\textit{AntiPattern Solution:} Industry partners are rewarded for product quality, the number of customers acquired and customer satisfaction, while researchers are rewarded for publications in high impact factor journals. In C1, for instance, for SRL researchers, one metric of interest was publications in high-impact journals and top conferences, while for industry partner Cisco the main metric has been maintaining high product quality through rigorous testing. Similarly, in C2, the main reward mechanism for SZ was high volume sales of their sensors and simplification of customer support, while for their academic partners it was about high quality publications. 

\textit{Symptoms and Consequences:}
Mismatches in rewards mechanisms can create isolation of partners, as each partner tries to optimize metrics that interest their organization. 

\textit{Refactored Solution:}
Each partner is the quintessential frog in its own well, unless they understand the reward mechanisms of each other. For instance, it is important for a research organization to realize that companies can only thrive financially when they obtain sales of their products and services. 
One solution is to have researcher work part-time in a company to understand the reward mechanisms. Similarly, it is also worthwhile for an industry practitioner to spend some time in a research group and understand the complexity of publishing high quality scientific articles and its benefits for the society. In C1, we had an industrial PhD student from ABB robotics who spent some time in a research group and published several papers. In addition, he facilitated the political decisions to incorporate research artifacts produced in the collaboration for testing ABB's painting robots. A researcher at SRL pointed out that \textit{"This factor greatly streamlined the collaboration between SRL and ABB robotics."} Similarly, in C2, the collaboration between SZ and University of Oslo was better aligned after the University of Oslo included SZ in research grant proposals as a partner and a vendor, to help increase the sales for SZ. In C3, reward mechanisms are defined as KPIs as part of consortium agreement, before the commencement of the project. These KPIs are something all partners in the project agree to and see as feasible goals to attain, despite their own reward systems in their companies. These KPIs are often specified to be simple and easy to achieve and often amenable to interpretation. For instance, in InterQ, a KPI is to achieve 99\% data completeness in acquiring manufacturing data and what is most relevant to the project is demonstrating this goal. 

\textit{Benefits:} Matching reward mechanisms can make it easy to develop a win-win situation for collaborators in a project. A collaboration can flourish if the partners understand each others reward mechanisms and the need for it in the society. Ideally, companies that can gain an edge from research will have increased sales of their products and services, while researchers can create high impact from their research work.

\textit{Related Solutions:} P3, P8, AP13.

\subsection{AP11: Arranged Marriage} 

\textit{Synopsis:} Long-term IA collaboration projects need to welcome change in different aspects of the collaboration (leadership, resources), as an opportunity to improve defective parts and be able to co-create value.

\textit{Context:} National and European councils that fund scientific research set criteria that require the establishment of long-term co-creation projects between diverse stakeholders, many of who may not have ever worked together in the past. It is in the best political and financial interests of the research councils to ensure that the long-term projects succeed. Very much like arranged marriages in India.

\textit{Anti-Pattern Solution:} Whenever a governmental research council decides to allocate funds to a project involving academic and industry partners, a consortium agreement is signed that outlines how time will be spent by each partner and how the intellectual results will be exploited. It also outlines how conflicts will be managed and mitigated. 

\textit{Symptoms and Consequences:} The stakeholder with responsibility of running the project is often asked to keep the funding bodies and research council \emph{happy and satisfied} about the ongoing collaboration. This often can lead to phrasing results and outcomes of a project with a steady streak of \emph{positiveness}. The non-engagement of some partners and poor outcomes are not presented with full clarity, since it is in nobody's interest, especially when the money comes from public funding with no real ownership. Reviewers are appointed to the job of evaluating projects where their role becomes limited to going through a checklist, as they do not really have "skin in the game". Elected politicians may sometimes be seen as in charge of  setting priorities and  allocating grants. However, very often their own terms are shorter than the length of long term co-creation projects. It has become very common to smooth communication of results to hide bigger issues underlying a project. This often leads to tremendous loss of resources and public tax payer money over several years. It also leads to sub-optimal and forced relationships between stakeholders in some projects that lack the necessary level of synergy that can stand the test of time  and be beneficial to the society.

\textit{Refactored Solution:} Negative results should be given as much importance as positive results  in co-creation projects. When partnerships do not reach a mutually beneficial agreement it should be easy to replace partners in long running projects. Intellectual property agreements should ensure that partners are fairly compensated for their contributions in case they are replaced. This may be seen analog to a divorce in a marriage. Similarly, change of personnel and leadership should be taken into account in research projects when conflicts arise frequently and are not resolved. These collaborations should be treated similar to change in leadership in businesses, where a board elects a CEO. All in all, the funding bodies should learn to embrace both positive and negative results as their own and see the value in the honesty. In C1, SRL has had to part ways with Norwegian Toll Customs and FMC technologies midway in the project, as the companies' strategy did not align with the goals of the project. They did not see value in spending more time and effort in the project after a few years of effort. This was communicated to the Norwegian Research Council and a new partner, the Cancer Registry of Norway, was added to the project. In C2, most public grants were secured by the University of Oslo and the collaboration with SZ lasted for short periods of time, whenever it was relevant. Here, the person of contact or the role of leader in a project varied from depending on who was best suited to manage the collaboration. In C3, the criteria enforced to have at least three European countries in a project, which can be seen as a peace project that fosters collaboration, while countries are getting to know each other despite cultural differences. A good approach to a successful collaboration is to have a solid foundation of partners who have worked with each other, and then add new partners in the project. For instance, SINTEF, IDEKO, DANOBAT, and TEKNIKER as a group and Renault, Predict and Comau as another group have previously worked together. While INLECOM, DNVGL, and TU Darmstadt are newcomers to the consortium. Despite the strong collaborative foundation, some of the new members have had reservations for the collaboration, due to potential conflicts of interest in intellectual property and in business partnerships. We mitigate the effect of such reservations in collaboration by creating separate groups of industrial partners in conflict of interest and neutral researchers who ensure that project results are equally shared by both parties.

\textit{Benefits:} There are many benefits of being flexible with agreements in long-term collaboration projects. When there is a lack of synergy, then partners can leave and be replaced. As a project manager at SZ reflected \textit{"It is in the best interest of all parties to say when things are not working, instead of hiding it, so that we can focus energy on what is working"}. Similarly, leadership of a publicly funded project should be mutable like in businesses. This can inspire honesty in projects and hope for change when projects are not running optimally and need  a fresh breath of energy. The flexibility of course comes at a cost that needs to be clearly quantified before major structural changes are made. The funding bodies should take an active role in evaluating alternate paths when they see that projects are not bringing the benefits to the society they initially hoped for.

\textit{Related Solutions:} P1, AP4, AP9.

\subsection{AP12: Repackaging ideas} 

\textit{Synopsis:} Stakeholders repackage old ideas and concepts into terminology developed for new trends and hype.

\textit{Context:} New technologies find themselves on the Gartner hype cycle every year and are much discussed in news and media. Funding agencies prepare call texts based on what is up and coming. The technologies move quickly on the hype cycle and hence, it is necessary for stakeholders, both academia and industry, to repackage something they have been working on for years to the demands and pressure of new trends and terminology.

\textit{AntiPattern Solution:} Industrial partners want to stay relevant by staying abreast of and use the latest  technological innovations. In an IA co-creation process, industry partners often rely on researchers to repackage what is being done for several years using a dated technology into a new technology. This is mainly due to the lack of resources to experiment with new technology in the industry. For instance, industry would like to experiment with NoSQL databases instead of relational databases, to become scalable to more data and users. Researchers on the other hand repackage their old concepts with new terminology. For instance, as an industry practitioner in C1 said \textit{"Constraint programming has been around for a couple of decades, but in contemporary times a preferred term would be symbolic AI to stay relevant and possibly increase a chance of obtaining funding. But in essence, if constraint programming did not work in practice because of scalability issues, neither will symbolic AI"}. 

\textit{Symptoms and Consequences:} Repackaging old ideas into new terms and technologies creates the following issues: (a) new technologies may not be relevant to the problem at hand, (b) the use of new technology may only be superficial if the real advantages of using new technology is not understood, (c) new technologies and terms may not pass the baptism time and may die out too early, and (d) a lot of money and person-hours can be used in pursuit of repackaging and reselling old ideas in a new framework.

\textit{Refactored Solution:} It is often important to embrace the emergence of a new technological trend, as it is a culmination of what is made possible with growing computational power and storage for instance. Both academia and industry want to stay relevant and up to date in their respective domains.  Instead of repackaging old ideas using new terminology and technologies, we suggest that the stakeholders identify the core problem in a mathematical formalism that is independent of new trends and technology, as a first step. This will help the team think clearly without being blinded by hype. This is also a place where old and new concepts have many aspects in common. The implementation of a solution however can use new technologies and terminology. In C1, researchers in SRL stripped a complex problem into mathematical symbols and statements to understand the real complexity. For instance, the problem of test selection and prioritization of video conferencing software at Cisco was specified as a learning problem \cite{Marijan2019ALA}, as a first step, followed by the choice of technologies, such as TensorFlow or PyTorch for machine learning, at a larger scale. In C2, the problem of predicting respiratory minute ventilation from ribcage movement data was initially attempted by creating a biophysical model of the respiration. This had its limitations, as ribcage movement measurements were varying from person to person. Therefore, it was necessary to attempt to address the same problem with contemporary deep learning technology. Scientists in the co-creation project specified the same problem as a deep learning problem, based on data collected from different people. The deep learning model was eventually trained and tested using novel machine learning frameworks, such as TensorFlow and PyTorch, with better results. In C3, the idea of controlling the manufacturing process using data from observing is not new. However, the use of deep learning models is trendy and hence the idea of using deep neural networks was repackaged for tasks such as improving data quality in faulty sensors.

\textit{Benefits:} The simplification of a problem to its bare bones mathematical formulation makes it trend-,  newfangled-terminology- and technology-agnostic. This brings clarity to the co-creation process, and all stakeholders are aligned to the core problem, instead of placing a lot of focus on finding a new wrapper as a solution to a problem. Eventually, once the problem is well-defined, stakeholders can make an informed decision on the type of technology to choose, to best solve the problem. 

\textit{Related Solutions:} AP6.

\subsection{AP13: McNamara fallacy} 

\textit{Synopsis:} When decisions are made in a co-creation process, it is only quantifiable results, outcomes, and observations that matter.

\textit{Context:} A co-creation process between researchers and practitioners needs to be evaluated on a regular basis by a grant-giving body. For a project spanning several years, such evaluation often takes place annually, midway, and at the end.

\textit{AntiPattern Solution:} The evaluation of a co-creation project is often performed using selected metrics. These metrics include number of publications, citation index, number and monetary value of grants accepted by researchers. While industry practitioners are evaluated based on sales figures and monetary value of projects acquired in the case of consultancy firms. 

\textit{Symptoms and Consequences:} Robert McNamara was an MBA from Harvard University in 1939. He was the first president of the Ford Company from outside the Ford family, and he eventually became the US Secretary of Defense during the Vietnam war. McNamara during his time in Ford was known to select data points and ruthlessly optimize the efficiency, costs, and quality at Ford. He brought the same approach to the Vietnam war where his metric for success was \emph{body count}. This was a poor measure of how a war is progressing, because it reduced the deeply human process to a mere figure, known as the \textit{McNamara fallacy}. Similarly, co-creation between researchers and industry practitioners can be a deeply human process spanning several months to years. A researcher in C2 reflected \textit{"Measuring outcomes based on quantifiable metrics make us forget about the social and human process behind the success or failure, which is wrong, because this process can be incredibly insightful and valuable to learn from"}. Very little time is spent on understanding the human experience of running a project. The sole focus on academic productivity based on metrics was associated with poorer physical health, increased burnout, and reduced productivity \cite{hodge2020balancing}. 

\textit{Refactored Solution:} The human process of co-creation needs to be recognized for its beauty and depth and also for the costs incurred. The process of co-creation should be documented with narrative, because it is valuable and insightful to know that behind technological innovations there are humans with flaws and challenges. The writing and documentation of the process helps externalize the non-quantifiable aspects that lead to quantifiable outcomes, such as a publication or a software artifact. For instance, in C2, there is a need to create awareness about improvement in physical health through the focus on breathing using the Flow sensor developed by SZ. This was achieved through simpler popular science communication in a press note \footnote{https://www.sporttechie.com/sweetzpot-launches-
ow-breathing-sensor-athletes/}. The press note led to increased awareness about breathing as a metric that we often ignore
in sports. In C3, gaining trust between research institutes and commercial partners in the manufacturing industry is of paramount importance, because researchers are seen as neutral between two or more companies competing in the same sector. An important thing that is not quantifiable is demonstrating that researchers are successfully able to balance the trade-off between openness needed for research and protecting the intellectual property of individual commercial partners and their interests in the project.

\textit{Benefits:} Focusing on not only quantifiable results helps lay emphasis on the often challenging human process that results in a technological innovation. Knowledge of the human process can help improve co-creation and enhance the state of flow among collaboration participants. This can drastically mitigate burnout rates. The management of stakeholders by neutral parties such as research institutes is never quantified, but the ensuing trust developed during the project helps such research institutes be invited to new collaborations in future.

\textit{Related Solutions:} P3, P8.

\subsection{AP14: Not invented here} 

\textit{Synopsis:} Collaboration partners show bias against internalizing the knowledge that originates from a different field of expertise.

\textit{Context:} Collaborative projects between industry and academia are developed with an idea that each side of the collaboration has an expertise required for addressing the project challenges, but not possessed by the other side. For example, in C1, SRL researchers had a unique expertise in combinatorial test optimization, which Cisco engineers did not have, and which was needed for optimizing the testing of video-conferencing systems at Cisco. On the other hand, Cisco possessed video-conferencing systems, and had a deep knowledge of the challenges and constraints for testing such systems. 

\textit{AntiPattern Solution:} In a typical scenario of a collaborative project, industry and academia initially meet to discuss industry practice and identify the challenges that industry is looking to solve. Discussions lead to the identification of a set of requirements for the solution to be developed. Afterwards, these two teams part ways and start working on the project with sparse interaction and limited opportunity to update each other on the direction of their work. A few months forward, academia is ready to have industry deploy the research prototype they have developed in the lab. However, industry avoids to use the prototype, believing that if it has not been developed in-house, it is not as valuable.

\textit{Symptoms and Consequences:} Bias against external knowledge by industry is developed as a consequence of not having insights in the knowledge development process, understanding of the underlying research concepts, nor the access to decision making about technical choices during the prototype development. In C1, in the example of Cisco video-conferencing system testing, the initial research prototype developed by SRL researchers had a similar destination, because there were technological incompatibilities between the prototype and the testing toolchain where the prototypes should have been integrated. Other reasons for having "not invented here syndrome" could include intellectual property concerns, costly absorption due to a steep learning curve of the core research concepts, or simply risks associated with the concepts unproved in the real world.

\textit{Refactored Solution:} It is critical to establish co-creation as a form of collaboration, where industry and academia involve in a process of continuous interaction during all stages of collaboration, from problem definition to solution deployment. Co-creation is able to build mutual trust and partnership between collaborating parties that leaves no room for the notion of "external" knowledge. There is only one knowledge developed, and it is equally owned by everyone involved in the project. For example, in C1, a co-creation was established between SRL researchers and ABB engineers through several strategies. Researchers spent a lot of time at the industry site, working side by side the ABBs engineers. There was an industrial PhD student employed by ABB and supervised by a SRL researcher, which helped catalyze co-creation, as this person was able to connect knowledge from both teams and make them realize the value of the expertize of one another. In C2, one challenge for the startup SZ was to demonstrate that their invention was cheaper and better in accuracy compared to existing solutions. The UiO research group on obstructive sleep apnea, in fact, tested the Flow sensor with several other low cost sensors \cite{b48}, to demonstrate that the Flow sensor indeed could be used for clinical studies. This external validation in a clinical study helped overcome the bias of not-invented-here and positioned SZ's Flow sensor for medical health applications, going beyond the initial focus on sports. In C3, it was not easy for the research institutes to obtain access to the data from the manufacturers for AI research, as the benefits of using AI was not very immediate, although advocated initially. Access to a data stream required several months of presentations and convincing with experiments on publicly available datasets. Yet, the most useful application for the manufacturing sensor data turned out to be data profiling and the applications of AI models were only a bonus. The use of AI is not very widespread in manufacturing and it is hard for industry actors to pause a bit and try to comprehend the benefits, as production runs 24 hours a day. Renault for instance mills 1500 cylinder heads per day in the Valladolid factory in Spain and has to see how AI benefits their process while in operation. Making an effort to understand how other partner's expertize can improve one's processes is a key to mitigate the not-invented-here syndrome. 

\textit{Benefits:} More efficient use of time and resources, less reinvention and duplication. There is no negative attitude towards using the knowledge developed in the collaboration project. Instead, there is a strong sense of ownership of such knowledge, as it was created by a joint effort from industry and academia. A researcher in C1 mentioned that \textit{"At the beginning of collaboration, we often see a disinterest of practitioners for the ideas we present. But once we establish trust and start interacting frequently, this bias completely disappears".}

\textit{Related Solutions:} P4, P5, P7, AP6.

\section{Discussion}
In this section, we suggest how to use the patterns and anti-patterns, and discuss their implications for practice. Next, we draw a line between our patterns and anti-patterns and best practices for IA collaboration suggested by previous studies, relative to the set of IA collaboration challenges observed in our experience and identified by others. Finally, we discuss limitations and threats to validity of our findings.

\subsection{Implications for Practice}
Applying the patterns and avoiding the anti-patterns provided in this article has shown to improve the success of IA collaborations, as observed across three projects in SE (see Section \ref{Contexts}). We believe that reusing these patterns in another IA collaboration project in SE will have a positive effect on the course of collaboration, especially if the patterns and anti-patterns are adapted to the collaboration context at hand. For example, in \textit{P4: Active dialog}, as a general advice we recommend to foster active dialog between industry and academia, which entails meeting and discussing regularly, and following up on the agreed actions in the agreed time-frame. However, we expect that the optimal frequency and volume of interaction will be different for different projects, and is, therefore, up for adjustment by individual participants.

The patterns and anti-patterns provided in this article are recommended and not prescriptive. The more of the patterns applied and anti-patterns avoided, the lower the risk of IA collaboration challenges, and the higher the probability of a successful collaboration. The lack of the patterns does not mean failure, nor applying them does guarantee success. It may be that even after applying the patterns and avoiding the anti-patterns the collaboration still fails. For example, due to the unstable financial situation, a company participating in an IA collaboration may need to prioritize other projects and pull resources out of the IA collaboration project, despite the established reciprocity between industry and academia, synergy in place, functioning active dialog and clear goals for the collaboration.

The set of 28 patterns and anti-patterns is not complete; it only captures our 14-year experience collected in three IA collaboration projects. Had we continued observing some of these projects further or started observing new collaboration projects, the set of patterns and anti-patterns might have been extended.   

Furthermore, among the 28 patterns and anti-patterns presented in this article, there are some which are more general and thus applicable to a wider range of collaboration contexts and others which are more context-dependent. For example, a more general pattern is \textit{P1: Reciprocity between stakeholders}, which entails that all participants must both give and receive in a collaborative project, for the project to be successful. We strongly believe that without the practice of reciprocity any collaboration will be derailed, not just the one in SE. On the other hand, \textit{P2: Standardized data exchange}, which promotes enabling the reuse of data, algorithms, and software code, is a pattern more relevant for the SE domain, and less for non-computer science domains. Other such SE-relevant patterns and anti-patterns may be \textit{P6: Minimum viable tools}, \textit{P11: Deriving research and business critical questions comes first and data sharing later}, \textit{P13: Anticipate unavailability of data}, \textit{AP2: Non-reusable minimum viable solutions}, \textit{AP8: Confound lab setup with the real world}.

The risk of applying the patterns wrongly or applying them when they do not fit is minimal since they are presented in a format conducive to reproducibility. The pattern description includes the context, explaining the circumstances under which the problem occurs, and the solution, explaining how the problem may be solved within that context. The anti-pattern description includes the symptoms, useful for recognizing a problematic solution, and consequences, explaining the implications of the problematic solution. If still the patterns and anti-patterns are used wrongly, the risk will be similar as if they were not used.

Finally, the presented patterns and anti-patterns are equally representative of both practitioners' and researchers' experience of IA collaboration. In Table \ref{tab:origin} we show which of the patterns and anti-patterns were initially mentioned by researchers and which by practitioners. However, all the patterns and anti-patterns were discussed with both researchers and practitioners, in order to develop the initially-collected findings into the final set of 28 solutions presented in this paper.

\begin{table}[ht]
\centering
\caption{Origin of patterns and anti-patterns.}
\begin{tabular}[t]{l>{\raggedright\arraybackslash}p{0.75\linewidth}}
\toprule
Initially mentioned by |&\textbf{Solution}\\
\midrule
Practitioners& P1, P2, P3, P4, P5, P6, P7, P9, P10, P11, P12, AP1, AP5, AP6, AP7, AP8, AP9, AP11, AP12 \\
Researchers& P1, P3, P4, P5, P7, P8, P9, P13, P14, AP1, AP2, AP3, AP4, AP5, AP6, AP7, AP10, AP13, AP14 \\
\bottomrule
\end{tabular}
\label{tab:origin}
\end{table}%

\subsection{Relation to Existing Evidence}
In 2016, Garousi \cite{b10} reviewed 33 studies discussing IA collaborations in SE, and synthesized a set of 127 best practices addressing 63 challenges identified across these studies. In Table \ref{tab:long} we map our patters (P) and anti-patterns (AP) to the challenges summarized in \cite{b10} and other challenges identified in our experience (denoted with \text{*} mark in front of the challenge), as well as previously suggested related best practices synthesized in \cite{b10}. If we have not observed a specific challenge in our experience, and thus do not provide a solution, this is shown as \ding{55} in the column "Our solutions". If there are no best practices addressing a related challenge identified in \cite{b10}, this is shown as \ding{55} in the column "Comparison with other solutions".

\begin{center}
\begin{longtable}{L{4cm} L{2cm} L{8cm}} 
\caption{Mapping of patterns and anti-patterns to different challenges and previous solutions for IA collaborations. \ding{69} denotes a novel challenge identified by us. \ding{55} denotes either a challenge we have not observed in our experience (2nd column) or that no best practices have been mapped to the specific challenge in \cite{b10} (3rd column).} \label{tab:long} \\

\hline \multicolumn{1}{L{4cm}}{\textbf{Challenges} (adapted from \cite{b10} and identified by us \ding{69})} & \multicolumn{1}{L{2cm}}{\textbf{Our solutions}} & \multicolumn{1}{L{8cm}}{\textbf{Comparison with other solutions} (synthesized in \cite{b10})} \\ \hline 
\endfirsthead

\multicolumn{3}{c}%
{{\bfseries \tablename\ \thetable{} -- continued from previous page}} \\
\hline \multicolumn{1}{L{4cm}}{\textbf{Challenges} (adapted from \cite{b10} and identified by us \ding{69})} & \multicolumn{1}{L{2cm}}{\textbf{Our solutions}} & \multicolumn{1}{L{8cm}}{\textbf{Comparison with other solutions} (synthesized in \cite{b10})} \\ \hline 
\endhead

\hline \multicolumn{3}{|r|}{{Continued on next page}} \\ \hline
\endfoot

\hline \hline
\endlastfoot

Research results are not relevant for practice & P8, P11 &  We suggest making a two-faceted problem definition, i.e. starting from a practical problem and deriving a research problem from it, as well as to focus on business critical questions at the beginning of collaboration. Whereas others suggest to make long-term commitments, provide frequent access for researchers, work in a team, use a use case study method, and pilot a solution with industry practitioners. \\
Research results are not measurable and exploitable  & P3, AP7 & \ding{55} \\
Researchers do not understand relevant problems from an industry point of view & P4 & We suggest building and maintaining active dialog throughout the collaboration to understand as much of the domain knowledge as possible, by means of workshops and seminars, which is in line with solutions proposed by others.   
\\ 
University education not focused on industrial
needs & \ding{55} & \ding{55}  \\
Research topic selection not driven by relevance & P8, P11 & \ding{55}  \\
Validity of research not properly addressed & P6, AP8 & \ding{55} \\
Running a flexible research project is challenging & P12, AP11 & \ding{55}  \\
Research in its nature is risky & AP9 & \ding{55} \\
Difficult to assess if research addresses future industry needs making it challenging to decide on solution & P7 & \ding{55}  \\
Integrating new/improved solutions in existing context & P5, P7 & \ding{55} \\
Deficiencies in software engineering education & \ding{55} & While we have not encountered this challenge in our experience, others suggest to work as a team, to make long-term commitment such that industry gets involved in education, and to use a use case study method for spreading knowledge.  \\
Lack of training, experience, and skills & \ding{55} & \ding{55}  \\
Deficiencies in research skills for practitioners & P4, P7 & \ding{55} \\
Lack of commitment and difficultly to assess research results and forums & P9 & \ding{55}  \\
Deficiencies in domain knowledge for researchers & P4, P7 & We suggest building an active dialog at all stages of collaboration using physical and virtual channels of interaction, as well as ensuring frequent interaction to demonstrate progress and intermediate results of technical developments often. Whereas others suggest to co-locate researchers on industry sites, to use established guidelines and data collection methods, and to employ researchers.  \\  
Lack of commitment to invest money & \ding{55} & \ding{55} \\
Lack of commitment to provide access and time & P1, AP4, AP5 & We suggest to identify the motivations and needs of all participants and develop a norm for reciprocity between stakeholders, as well as to ensure that all participants have "skin in the game". Whereas others focus on champions, making long-term commitments, proper presentation by researchers in early meetings, proper topic selection, showing benefits of the research solution for the industrial partner, collocating researchers on industry side, ensuring frequent interaction through meetings, managing intellectual property rights, and piloting the solution with industry practitioners.  \\
Lack of commitment due to human factors & AP14 & \ding{55}  \\
Lack of commitment due to competitive business & P1, AP4 & \ding{55}  \\
Different time horizons for industry and academia  & P14 & \ding{55} \\
Different interests and objectives & P4, P8 & \ding{55}  \\ 
Different perception of what solutions are useful & P4, AP1, AP13 & \ding{55} \\
Different terminology and ways of communicating &  AP6 & We suggest making an effort from early stages of collaboration to understand and synchronize terminology, through day-to-day interaction and lightning talks on specific topic. Whereas others suggest having prior positive experience to facilitate communication, as well as to personally interact with practitioners during data collection. \\
Different reward systems & AP1, AP10 & \ding{55}  \\
Different communication channels and directions of information flow & P4, AP6 & We suggest building an active dialog between industry and academia at all stages of collaboration, while focusing on a common vocabulary. Whereas others focus on workshops and seminars to increase visibility show relevance, strength and ability.  \\
Different cultures & P4, P8, AP1 & \ding{55} \\
Different expectations on quality of evidence
in research & P6, AP3 & \ding{55}  \\
Different focus on scale of solutions & AP2, AP3, AP8 & \ding{55} \\
Different types of knowledge available & P5, P7 & \ding{55}  \\
Technology push from academia greater than technology pull from industry & AP14 & \ding{55} \\
Different contexts & P4, P8, P9 & \ding{55} \\
Different business models & \ding{55} & \ding{55} \\
Different perception of challenges & P4, P8, AP8 & \ding{55} \\
Different requirements on novelty & P14, AP3, AP12 & \ding{55} \\
Communication gaps between researchers and
practitioners & AP6 & \ding{55} \\
Difficulty of managing multiple research partners & \ding{55} & \ding{55} \\
Difficulty to elicit information from developers & P4, P7 & \ding{55} \\
Communicating on time-frames, topics, and responsibilities & P4, P7 & \ding{55} \\
Lack of prior relationships between a company
and academia & \ding{55} & \ding{55} \\
Resistance to change and inflexibility & AP11, AP14 & We suggest to welcome change in different aspects of the collaboration as an opportunity to improve defective parts, as well as to embrace openness for knowledge that originates from a different field of expertise. Whereas other suggest to use the case study method. \\
Difficulties in training practitioners due to
high training cost and lack of time & \ding{55} & \ding{55} \\
Lack of organizational stability and continuity & \ding{55} & We have not seen this challenge in our experience, while others focus on ensuring management engagement on industry side. \\
Intangible human factors with organization-wide impact & AP13 & \ding{55} \\
Competition between industrial and external
researchers & P5, AP14 & \ding{55} \\
Hard to find champions & AP4 & \ding{55} \\    
Difficulty to achieve clear and realistic goals & P4, P7 & Our solutions include active dialog and frequent iteration to align goals, which is in accordance with solutions by others, which focus on proper presentation and communication by researchers in early meetings. \\	
Solution incompatible with organizational culture & P7 & \ding{55} \\
Lack of willingness to invest time/effort & P1, P6, AP4, AP5 & We suggest to develop a norm of reciprocity between partners, develop a minimum viable proof of concept in early stages of collaboration, to demonstrate practical benefits of research concepts, to have incurred risk for all partners by being involved in achieving a goal, and to discuss value creation with clarity in the start of a project. Solutions proposed by others, which include showing benefits of the research solution for the industrial partner, are in line with our experience. \\
Difficult to find the right project infrastructure
(management, collaboration environments) & \ding{55} & \ding{55} \\
Difficulty to integrate external competence & P7, P10, AP14 & \ding{55} \\
Time-critical windows of opportunity for product research & \ding{55} & \ding{55} \\
Lack of openness to disclose weaknesses & AP11 & \ding{55} \\
Loss of champions in projects & \ding{55} & \ding{55} \\
Lack of resources due to over-investment & \ding{55} & \ding{55} \\
Financial investment risky from academic side & \ding{55} & \ding{55} \\
Licensing restrictions on tools & P10 & \ding{55} \\
Lack of resources to provide technical support for research solutions & P10 & \ding{55} \\
Intellectual property rights and privacy access to data & P10, P13 & \ding{55} \\ 
Difficulty in managing intellectual property rights & P10 & \ding{55} \\
Missing trust and respect & P4, P7, P12, AP4 & We focus on active dialog and frequent iteration to build trust and align goals, as well as to minimize moving parts to increase project stability, and ensure everyone has incurred risk in the project. Whereas others suggest to establish common and simple terminology, show benefits of research solutions for the industrial partner, to work in a team, and personally interact with practitioners during data collection. \\
Incorporating new methods and solutions in
research contacts & \ding{55} & \ding{55} \\
\ding{69} Limited sharing of potentially reusable artifacts & P2 &  \ding{55}    \\
\ding{69} Lack of objective evidence showing the practical value of the knowledge created in early stages of a collaborative project & P6 & \ding{55}\\
\ding{69} Difficult to transfer agile practices from industry to  research knowledge creation & P14, AP3 & \ding{55} \\
\ding{69} Developing and testing research results in the lab vs real world & AP8 & \ding{55}\\
\ding{69} Keeping abreast of latest technology innovations & AP12 & \ding{55}\\ 

\end{longtable}
\end{center}

\subsection{Limitations and Threats to Validity}
\textbf{Construct Validity:} 
To reduce a potential threat to construct validity, during the interviews, each time we asked a question, we made sure that the interviewee understands the question and interprets it the same way as the interviewer. Another threat to construct validity is related to the observer-expectancy effect in the observation sessions, which means that the observer's presence influences the informants' behavior. To mitigate this threat, the observers acted as "normal" participants, taking notes on a laptop, trying to be as unobtrusive as possible. Another validity threat may be related to the accountability of the data collection period. During interviews, our respondents may have forgotten to mention an important aspect of IA collaboration, as it occurred long before the interviews were conducted. To mitigate this threat, during interviews, we did not rush the discussion, and after interviews, we encouraged the respondents to contact us later if they think of any feedback they would like to provide related to the interviews.

\textbf{External Validity:}
The reported patterns and anti-patterns are empirically tested in three different collaboration contexts: a large-scale long-term context, small-scale mid-term context, and large-scale mid-term context. They are not tested in a wide variety of industry-academia collaboration setups, which could reduce the generalizability of our findings. However, as mentioned by Neill \cite{Colin}, an important aspect of patterns and anti-patterns is the "rule-of-three", which means that patterns and anti-patterns must have been used successfully in practice three times, to be called patterns and anti-patterns. Furthermore, out patterns and anti-patterns have been tested in different countries (two of the tree collaboration contexts are international), in different organizations and different industries (software, hardware, manufacturing, public sector). To further improve the external validity of our findings, we will revisit, expand and improve the set of 28 patterns and anti-patterns based on collected experiences from the collaboration projects. Another threat to external validity is that our results may be less representative of the SE areas not covered by our three collaboration contexts. However, our studied collaboration contexts cover a wide range of SE areas, mitigating this particular threat. Furthermore, although the patterns and anti-patterns are derived from the three collaboration contexts studied, they address IA collaboration challenges found in other contexts (See Table \ref{tab:long}, and thus we believe the patterns and anti-patterns could be transferable to other contexts. However, the list of patterns and anti-patterns is not exhaustive. It is reflective of only the experiences we have from the three collaboration contexts in SE. Observing another collaboration context may lead to new patterns.

\textbf{Internal Validity:} Participant observation used for data collection has the limitation of being subjective, as it represents a perspective of an observer. To minimize this internal validity threat, we had two participant observers who collected data, and during data analysis, we applied observer triangulation, where conclusion made by one observer were checked by another. Furthermore, we applied data source triangulation, combining and relating data coming from different sources, to increase the validity of our empirical study. Next, our selected respondents may not represent all relevant participants in the collaboration. However, we conducted participant observation such to involve participants with different experience and expertise. For example, software engineers and researchers were observed as part of joint teams, and managers were observed in focus group discussions and workshops. In interviews, we applied purposive sampling, based on the observer's judgment, selecting participants who may provide the most useful information related to the questions that arose during the previous cycle of data analysis, as well as to ensure that both positive and negative aspects of the collaboration are captured. As there is a risk of the selected sample not being representative to the population, in some data collection cycles, we applied a maximum variation sampling, selecting interviewees such to vary their backgrounds and expertize. Some of the interview participants expressed negative experience of some phases of collaboration, for example, the lack of information flow between industry and academia. We captured such hurdles with special interest, probing deeper into the aspects that did not work, their effect on the collaboration, and possible ways of resolving the hurdles. The negative experience collected provided a valuable input for anti-pattern definition. Further factors that can limit the internal validity of our study are external factors that could affect the cause effect relationship between patterns and anti-patterns and results of co-creation. For example, the background of the researchers and practitioners, the timing of co-creation, the financial status of companies, market dynamics, churn rate in employees or force majeure such as corona epidemic, which changed the ways of interacting between industry and academia, from a real-world to virtual interaction.     

\textbf{Conclusion Validity:} To reduce the threat of reaching the wrong conclusions from the data, we triangulated using multiple sources of data, such as field notes from participant observation, findings from interviews, emails and feedback from the stakeholders about the derived patters and anti-patterns. We also applied observer triangulation, where two observers collected and analyzed data, while reviewing each other's findings, to reduce conclusion bias. Furthermore, feedback from the stakeholders was collected using a member checking technique, where the patterns and anti-patterns were presented to the stakeholders to get their opinion.

\section{Conclusion}
In this paper we discuss our experience of co-creation, as means to industry-academia collaboration, gained in three collaboration setups in the area of software engineering research. Throughout this experience, we have observed a set of 28 recurring best practices and issues to avoid, which we provide as the reported patterns and anti-patterns. We exemplify the patterns and anti-patterns using three different IA collaboration projects. Such exemplified insights into recurring patterns of successes and failures can positively contribute to other industry-academia collaboration projects in software engineering. 

\bibliographystyle{ACM-Reference-Format}
\bibliography{acmart}


\begin{thebibliography}{00}


\ifx \showCODEN    \undefined \def \showCODEN     #1{\unskip}     \fi
\ifx \showDOI      \undefined \def \showDOI       #1{#1}\fi
\ifx \showISBNx    \undefined \def \showISBNx     #1{\unskip}     \fi
\ifx \showISBNxiii \undefined \def \showISBNxiii  #1{\unskip}     \fi
\ifx \showISSN     \undefined \def \showISSN      #1{\unskip}     \fi
\ifx \showLCCN     \undefined \def \showLCCN      #1{\unskip}     \fi
\ifx \shownote     \undefined \def \shownote      #1{#1}          \fi
\ifx \showarticletitle \undefined \def \showarticletitle #1{#1}   \fi
\ifx \showURL      \undefined \def \showURL       {\relax}        \fi
\providecommand\bibfield[2]{#2}
\providecommand\bibinfo[2]{#2}
\providecommand\natexlab[1]{#1}
\providecommand\showeprint[2][]{arXiv:#2}

\bibitem[\protect\citeauthoryear{A.~Sandberg}{A.~Sandberg}{2011}]%
        {b15}
\bibfield{author}{\bibinfo{person}{T.~Arts A.~Sandberg, L.~Pareto}.}
  \bibinfo{year}{2011}\natexlab{}.
\newblock \showarticletitle{Agile Collaborative Research: Action Principles for
  Industry-Academia Collaboration}.
\newblock \bibinfo{journal}{{\em IEEE Software\/}} \bibinfo{volume}{28},
  \bibinfo{number}{04} (\bibinfo{date}{jul} \bibinfo{year}{2011}),
  \bibinfo{pages}{74--83}.
\newblock
\showISSN{1937-4194}
\showDOI{%
\url{https://doi.org/10.1109/MS.2011.49}}


\bibitem[\protect\citeauthoryear{Avison, Lau, Myers, and Nielsen}{Avison
  et~al\mbox{.}}{1999}]%
        {Avison}
\bibfield{author}{\bibinfo{person}{David~E. Avison}, \bibinfo{person}{Francis
  Lau}, \bibinfo{person}{Michael~D. Myers}, {and} \bibinfo{person}{Peter~Axel
  Nielsen}.} \bibinfo{year}{1999}\natexlab{}.
\newblock \showarticletitle{Action Research}.
\newblock \bibinfo{journal}{{\em Commun. ACM\/}} \bibinfo{volume}{42},
  \bibinfo{number}{1} (\bibinfo{year}{1999}), \bibinfo{pages}{94--97}.
\newblock


\bibitem[\protect\citeauthoryear{Barroca, Sharp, Salah, Taylor, and
  Gregory}{Barroca et~al\mbox{.}}{2018}]%
        {b29}
\bibfield{author}{\bibinfo{person}{Leonor Barroca}, \bibinfo{person}{Helen
  Sharp}, \bibinfo{person}{Dina Salah}, \bibinfo{person}{Katie Taylor}, {and}
  \bibinfo{person}{Peggy Gregory}.} \bibinfo{year}{2018}\natexlab{}.
\newblock \showarticletitle{Bridging the gap between research and agile
  practice: an evolutionary model}.
\newblock \bibinfo{journal}{{\em International Journal of System Assurance
  Engineering and Management\/}} \bibinfo{volume}{9}, \bibinfo{number}{2}
  (\bibinfo{date}{01 Apr} \bibinfo{year}{2018}), \bibinfo{pages}{323--334}.
\newblock
\showISSN{0976-4348}
\showDOI{%
\url{https://doi.org/10.1007/s13198-015-0355-5}}


\bibitem[\protect\citeauthoryear{{Basili}, {Briand}, {Bianculli}, {Nejati},
  {Pastore}, and {Sabetzadeh}}{{Basili} et~al\mbox{.}}{2018}]%
        {b2}
\bibfield{author}{\bibinfo{person}{V. {Basili}}, \bibinfo{person}{L. {Briand}},
  \bibinfo{person}{D. {Bianculli}}, \bibinfo{person}{S. {Nejati}},
  \bibinfo{person}{F. {Pastore}}, {and} \bibinfo{person}{M. {Sabetzadeh}}.}
  \bibinfo{year}{2018}\natexlab{}.
\newblock \showarticletitle{Software Engineering Research and Industry: A
  Symbiotic Relationship to Foster Impact}.
\newblock \bibinfo{journal}{{\em IEEE Software\/}} \bibinfo{volume}{35},
  \bibinfo{number}{5} (\bibinfo{year}{2018}), \bibinfo{pages}{44--49}.
\newblock
\showDOI{%
\url{https://doi.org/10.1109/MS.2018.290110216}}


\bibitem[\protect\citeauthoryear{Bosch}{Bosch}{2014}]%
        {b6}
\bibfield{author}{\bibinfo{person}{Jan Bosch}.}
  \bibinfo{year}{2014}\natexlab{}.
\newblock \bibinfo{booktitle}{{\em Continuous Software Engineering: An
  Introduction}}.
\newblock \bibinfo{publisher}{Springer International Publishing},
  \bibinfo{address}{Cham}, \bibinfo{pages}{3--13}.
\newblock
\showISBNx{978-3-319-11283-1}


\bibitem[\protect\citeauthoryear{Bradley, Hayter, and Link}{Bradley
  et~al\mbox{.}}{2013}]%
        {b31}
\bibfield{author}{\bibinfo{person}{Samantha~R. Bradley},
  \bibinfo{person}{Christopher~S. Hayter}, {and} \bibinfo{person}{Albert~N.
  Link}.} \bibinfo{year}{2013}\natexlab{}.
\newblock \bibinfo{booktitle}{{\em {Models and Methods of University Technology
  Transfer}}}.
\newblock \bibinfo{type}{UNCG Economics Working Papers} 13-10.
  \bibinfo{institution}{University of North Carolina at Greensboro, Department
  of Economics}.
\newblock
\showURL{%
\url{https://ideas.repec.org/p/ris/uncgec/2013_010.html}}


\bibitem[\protect\citeauthoryear{{Briand}}{{Briand}}{2012}]%
        {b1}
\bibfield{author}{\bibinfo{person}{L. {Briand}}.}
  \bibinfo{year}{2012}\natexlab{}.
\newblock \showarticletitle{Embracing the Engineering Side of Software
  Engineering}.
\newblock \bibinfo{journal}{{\em IEEE Software\/}} \bibinfo{volume}{29},
  \bibinfo{number}{4} (\bibinfo{year}{2012}), \bibinfo{pages}{96--96}.
\newblock
\showDOI{%
\url{https://doi.org/10.1109/MS.2012.86}}


\bibitem[\protect\citeauthoryear{{Briand}, {Bianculli}, {Nejati}, {Pastore},
  and {Sabetzadeh}}{{Briand} et~al\mbox{.}}{2017}]%
        {b3}
\bibfield{author}{\bibinfo{person}{L. {Briand}}, \bibinfo{person}{D.
  {Bianculli}}, \bibinfo{person}{S. {Nejati}}, \bibinfo{person}{F. {Pastore}},
  {and} \bibinfo{person}{M. {Sabetzadeh}}.} \bibinfo{year}{2017}\natexlab{}.
\newblock \showarticletitle{The Case for Context-Driven Software Engineering
  Research: Generalizability Is Overrated}.
\newblock \bibinfo{journal}{{\em IEEE Software\/}} \bibinfo{volume}{34},
  \bibinfo{number}{5} (\bibinfo{year}{2017}), \bibinfo{pages}{72--75}.
\newblock
\showDOI{%
\url{https://doi.org/10.1109/MS.2017.3571562}}


\bibitem[\protect\citeauthoryear{{Briand}}{{Briand}}{2011}]%
        {b43}
\bibfield{author}{\bibinfo{person}{L.~C. {Briand}}.}
  \bibinfo{year}{2011}\natexlab{}.
\newblock \showarticletitle{Useful software engineering research - leading a
  double-agent life}. In \bibinfo{booktitle}{{\em 2011 27th IEEE International
  Conference on Software Maintenance (ICSM)}}. \bibinfo{pages}{2--2}.
\newblock
\showDOI{%
\url{https://doi.org/10.1109/ICSM.2011.6080766}}


\bibitem[\protect\citeauthoryear{Brown, Malveau, McCormick~III, and
  Mowbray}{Brown et~al\mbox{.}}{1998}]%
        {b36}
\bibfield{author}{\bibinfo{person}{W.J. Brown}, \bibinfo{person}{R.C. Malveau},
  \bibinfo{person}{H.W. McCormick~III}, {and} \bibinfo{person}{T.J. Mowbray}.}
  \bibinfo{year}{1998}\natexlab{}.
\newblock \bibinfo{booktitle}{{\em AntiPatterns, Refactoring Software,
  Architectures and Projects in Crisis}}.
\newblock \bibinfo{publisher}{Wiley Computer Publishing}.
\newblock


\bibitem[\protect\citeauthoryear{Chimalakonda, Reddy, and Shukla}{Chimalakonda
  et~al\mbox{.}}{2015}]%
        {b7}
\bibfield{author}{\bibinfo{person}{Sridhar Chimalakonda},
  \bibinfo{person}{Y.~Raghu Reddy}, {and} \bibinfo{person}{Rakesh Shukla}.}
  \bibinfo{year}{2015}\natexlab{}.
\newblock \showarticletitle{Moving Beyond: Insights from 1st International
  Workshop on Software Engineering Research and Industrial Practices (SER-IPs
  2014)}.
\newblock \bibinfo{journal}{{\em SIGSOFT Softw. Eng. Notes\/}}
  \bibinfo{volume}{40}, \bibinfo{number}{2} (\bibinfo{date}{April}
  \bibinfo{year}{2015}), \bibinfo{pages}{28--31}.
\newblock
\showISSN{0163-5948}
\showDOI{%
\url{https://doi.org/10.1145/2735399.2735418}}


\bibitem[\protect\citeauthoryear{{Connor}, {Buchan}, and {Petrova}}{{Connor}
  et~al\mbox{.}}{2009}]%
        {b40}
\bibfield{author}{\bibinfo{person}{A.~M. {Connor}}, \bibinfo{person}{J.
  {Buchan}}, {and} \bibinfo{person}{K. {Petrova}}.}
  \bibinfo{year}{2009}\natexlab{}.
\newblock \showarticletitle{Bridging the Research-Practice Gap in Requirements
  Engineering through Effective Teaching and Peer Learning}. In
  \bibinfo{booktitle}{{\em 2009 Sixth International Conference on Information
  Technology: New Generations}}. \bibinfo{pages}{678--683}.
\newblock
\showDOI{%
\url{https://doi.org/10.1109/ITNG.2009.134}}


\bibitem[\protect\citeauthoryear{C.Wohlin}{C.Wohlin}{2013}]%
        {b13}
\bibfield{author}{\bibinfo{person}{C.Wohlin}.} \bibinfo{year}{2013}\natexlab{}.
\newblock \showarticletitle{Software engineering research under the lamppost}.
\newblock \bibinfo{journal}{{\em International Joint Conference on Software
  Technologies\/}} (\bibinfo{year}{2013}).
\newblock


\bibitem[\protect\citeauthoryear{Emerson, Fretz, and Shaw}{Emerson
  et~al\mbox{.}}{2001}]%
        {b33}
\bibfield{author}{\bibinfo{person}{R. Emerson}, \bibinfo{person}{R. Fretz},
  {and} \bibinfo{person}{L. Shaw}.} \bibinfo{year}{2001}\natexlab{}.
\newblock \bibinfo{booktitle}{{\em Participant observation and fieldnotes,
  Handbook of Ethnography}}.
\newblock


\bibitem[\protect\citeauthoryear{Falvagno and Dalli}{Falvagno and
  Dalli}{2014}]%
        {b25}
\bibfield{author}{\bibinfo{person}{M. Falvagno} {and} \bibinfo{person}{D.
  Dalli}.} \bibinfo{year}{2014}\natexlab{}.
\newblock \showarticletitle{Theory of value co-creation: a systematic
  literature review}.
\newblock \bibinfo{journal}{{\em Managing Service Quality\/}}
  \bibinfo{volume}{24}, \bibinfo{number}{6} (\bibinfo{year}{2014}).
\newblock


\bibitem[\protect\citeauthoryear{Gamma, Helm, Johnson, and Vlissides}{Gamma
  et~al\mbox{.}}{1994}]%
        {b37}
\bibfield{author}{\bibinfo{person}{E. Gamma}, \bibinfo{person}{R. Helm},
  \bibinfo{person}{R. Johnson}, {and} \bibinfo{person}{J. Vlissides}.}
  \bibinfo{year}{1994}\natexlab{}.
\newblock \bibinfo{booktitle}{{\em Design Patterns: Elements of Reusable
  Object-Oriented Software}}.
\newblock \bibinfo{publisher}{Addison-Wesley}.
\newblock


\bibitem[\protect\citeauthoryear{Garousi, Eskandar, and Herkilo\u{g}lu}{Garousi
  et~al\mbox{.}}{2017a}]%
        {Garousi2016}
\bibfield{author}{\bibinfo{person}{Vahid Garousi}, \bibinfo{person}{Matt~M.
  Eskandar}, {and} \bibinfo{person}{Kadir Herkilo\u{g}lu}.}
  \bibinfo{year}{2017}\natexlab{a}.
\newblock \showarticletitle{Industry-Academia Collaborations in Software
  Testing: Experience and Success Stories from Canada and Turkey}.
\newblock \bibinfo{journal}{{\em Software Quality Journal\/}}
  \bibinfo{volume}{25}, \bibinfo{number}{4} (\bibinfo{year}{2017}),
  \bibinfo{pages}{1091--1143}.
\newblock


\bibitem[\protect\citeauthoryear{Garousi, Felderer, Fernandes, Pfahl, and
  M\"{a}ntyl\"{a}}{Garousi et~al\mbox{.}}{2017b}]%
        {b11}
\bibfield{author}{\bibinfo{person}{Vahid Garousi}, \bibinfo{person}{Michael
  Felderer}, \bibinfo{person}{Jo\~{a}o~M. Fernandes}, \bibinfo{person}{Dietmar
  Pfahl}, {and} \bibinfo{person}{Mika~V. M\"{a}ntyl\"{a}}.}
  \bibinfo{year}{2017}\natexlab{b}.
\newblock \showarticletitle{Industry-Academia Collaborations in Software
  Engineering: An Empirical Analysis of Challenges, Patterns and Anti-Patterns
  in Research Projects}. In \bibinfo{booktitle}{{\em Proceedings of the 21st
  International Conference on Evaluation and Assessment in Software
  Engineering}} {\em (\bibinfo{series}{EASE'17})}. \bibinfo{pages}{224--229}.
\newblock
\showISBNx{9781450348041}
\showDOI{%
\url{https://doi.org/10.1145/3084226.3084279}}


\bibitem[\protect\citeauthoryear{Garousi, Petersen, and Ozkan}{Garousi
  et~al\mbox{.}}{2016}]%
        {b10}
\bibfield{author}{\bibinfo{person}{Vahid Garousi}, \bibinfo{person}{Kai
  Petersen}, {and} \bibinfo{person}{Baris Ozkan}.}
  \bibinfo{year}{2016}\natexlab{}.
\newblock \showarticletitle{Challenges and best practices in industry-academia
  collaborations in software engineering: A systematic literature review}.
\newblock \bibinfo{journal}{{\em Information and Software Technology\/}}
  \bibinfo{volume}{79} (\bibinfo{year}{2016}), \bibinfo{pages}{106--127}.
\newblock
\showISSN{0950-5849}
\showDOI{%
\url{https://doi.org/10.1016/j.infsof.2016.07.006}}


\bibitem[\protect\citeauthoryear{{Garousi}, {Pfahl}, and {Fernandes}}{{Garousi}
  et~al\mbox{.}}{2019}]%
        {Garousi2019}
\bibfield{author}{\bibinfo{person}{V. {Garousi}}, \bibinfo{person}{D. {Pfahl}},
  {and} \bibinfo{person}{J. {Fernandes}}.} \bibinfo{year}{2019}\natexlab{}.
\newblock \showarticletitle{Characterizing industry-academia collaborations in
  software engineering: evidence from 101 projects}.
\newblock \bibinfo{journal}{{\em Empirical Software Engineering\/}}
  \bibinfo{volume}{24} (\bibinfo{year}{2019}), \bibinfo{pages}{2540--2602}.
\newblock


\bibitem[\protect\citeauthoryear{Garousi, Shepherd, and Herkiloglu}{Garousi
  et~al\mbox{.}}{2020}]%
        {Garousi26}
\bibfield{author}{\bibinfo{person}{Vahid Garousi}, \bibinfo{person}{David~C.
  Shepherd}, {and} \bibinfo{person}{Kadir Herkiloglu}.}
  \bibinfo{year}{2020}\natexlab{}.
\newblock \showarticletitle{Successful Engagement of Practitioners and Software
  Engineering Researchers: Evidence From 26 International Industry-Academia
  Collaborative Projects}.
\newblock \bibinfo{journal}{{\em IEEE Software\/}} \bibinfo{volume}{37},
  \bibinfo{number}{6} (\bibinfo{year}{2020}), \bibinfo{pages}{65--75}.
\newblock
\showDOI{%
\url{https://doi.org/10.1109/MS.2019.2914663}}


\bibitem[\protect\citeauthoryear{{Gislason Bern}}{{Gislason Bern}}{2018}]%
        {b27}
\bibfield{author}{\bibinfo{person}{B. {Gislason Bern}}.}
  \bibinfo{year}{2018}\natexlab{}.
\newblock \showarticletitle{From Theory to Practice: Experiences of
  Industry-Academia Collaboration from a Practitioner}. In
  \bibinfo{booktitle}{{\em 2018 IEEE/ACM 5th International Workshop on Software
  Engineering Research and Industrial Practice (SER IP)}}.
  \bibinfo{pages}{22--23}.
\newblock


\bibitem[\protect\citeauthoryear{Gorschek, Garre, Larsson, and Wohlin}{Gorschek
  et~al\mbox{.}}{2006a}]%
        {Gorschek}
\bibfield{author}{\bibinfo{person}{Tony Gorschek}, \bibinfo{person}{Per Garre},
  \bibinfo{person}{Stig Larsson}, {and} \bibinfo{person}{Claes Wohlin}.}
  \bibinfo{year}{2006}\natexlab{a}.
\newblock \showarticletitle{A Model for Technology Transfer in Practice}.
\newblock \bibinfo{journal}{{\em IEEE Software\/}} \bibinfo{volume}{23},
  \bibinfo{number}{6} (\bibinfo{year}{2006}), \bibinfo{pages}{88--95}.
\newblock
\showDOI{%
\url{https://doi.org/10.1109/MS.2006.147}}


\bibitem[\protect\citeauthoryear{Gorschek, Garre, Larsson, and Wohlin}{Gorschek
  et~al\mbox{.}}{2006b}]%
        {b19}
\bibfield{author}{\bibinfo{person}{Tony Gorschek}, \bibinfo{person}{Per Garre},
  \bibinfo{person}{Stig Larsson}, {and} \bibinfo{person}{Claes Wohlin}.}
  \bibinfo{year}{2006}\natexlab{b}.
\newblock \showarticletitle{A Model for Technology Transfer in Practice}.
\newblock \bibinfo{journal}{{\em IEEE Software\/}} \bibinfo{volume}{23},
  \bibinfo{number}{6} (\bibinfo{date}{Nov.} \bibinfo{year}{2006}),
  \bibinfo{pages}{88--95}.
\newblock
\showISSN{0740-7459}
\showDOI{%
\url{https://doi.org/10.1109/MS.2006.147}}


\bibitem[\protect\citeauthoryear{Hodge, Wright, and Bennett}{Hodge
  et~al\mbox{.}}{2020}]%
        {hodge2020balancing}
\bibfield{author}{\bibinfo{person}{Brad Hodge}, \bibinfo{person}{Brad Wright},
  {and} \bibinfo{person}{Pauleen Bennett}.} \bibinfo{year}{2020}\natexlab{}.
\newblock \showarticletitle{Balancing effort and rewards at university:
  Implications for physical health, mental health, and academic outcomes}.
\newblock \bibinfo{journal}{{\em Psychological reports\/}}
  \bibinfo{volume}{123}, \bibinfo{number}{4} (\bibinfo{year}{2020}),
  \bibinfo{pages}{1240--1259}.
\newblock


\bibitem[\protect\citeauthoryear{Ind and Coates}{Ind and Coates}{2013}]%
        {b26}
\bibfield{author}{\bibinfo{person}{N. Ind} {and} \bibinfo{person}{N. Coates}.}
  \bibinfo{year}{2013}\natexlab{}.
\newblock \showarticletitle{The meanings of creation}.
\newblock \bibinfo{journal}{{\em European Business Review\/}}
  (\bibinfo{year}{2013}).
\newblock


\bibitem[\protect\citeauthoryear{Ivanov, Rogers, Succi, Yi, and Zorin}{Ivanov
  et~al\mbox{.}}{2017}]%
        {b9}
\bibfield{author}{\bibinfo{person}{Vladimir Ivanov}, \bibinfo{person}{Alan
  Rogers}, \bibinfo{person}{Giancarlo Succi}, \bibinfo{person}{Jooyong Yi},
  {and} \bibinfo{person}{Vasilii Zorin}.} \bibinfo{year}{2017}\natexlab{}.
\newblock \showarticletitle{What Do Software Engineers Care about? Gaps between
  Research and Practice}. In \bibinfo{booktitle}{{\em Proceedings of the 2017
  11th Joint Meeting on Foundations of Software Engineering}} {\em
  (\bibinfo{series}{ESEC/FSE 2017})}. \bibinfo{pages}{890--895}.
\newblock
\showISBNx{9781450351058}
\showDOI{%
\url{https://doi.org/10.1145/3106237.3117778}}


\bibitem[\protect\citeauthoryear{{J. Neill}, {A. Laplante}, and {F.
  DeFranco}}{{J. Neill} et~al\mbox{.}}{2012}]%
        {Colin}
\bibfield{author}{\bibinfo{person}{Colin {J. Neill}}, \bibinfo{person}{Philip
  {A. Laplante}}, {and} \bibinfo{person}{Joanna {F. DeFranco}}.}
  \bibinfo{year}{2012}\natexlab{}.
\newblock \bibinfo{booktitle}{{\em Antipatterns: Managing Software
  Organizations and People, Second Edition}}.
\newblock \bibinfo{publisher}{Taylor and Francis}.
\newblock
\showISBNx{978-1439861868}


\bibitem[\protect\citeauthoryear{{Jain}, {Ali Babar}, and {Fernandez}}{{Jain}
  et~al\mbox{.}}{2013}]%
        {b42}
\bibfield{author}{\bibinfo{person}{S. {Jain}}, \bibinfo{person}{M. {Ali
  Babar}}, {and} \bibinfo{person}{J. {Fernandez}}.}
  \bibinfo{year}{2013}\natexlab{}.
\newblock \showarticletitle{Conducting empirical studies in industry: Balancing
  rigor and relevance}. In \bibinfo{booktitle}{{\em 2013 1st International
  Workshop on Conducting Empirical Studies in Industry (CESI)}}.
  \bibinfo{pages}{9--14}.
\newblock
\showDOI{%
\url{https://doi.org/10.1109/CESI.2013.6618463}}


\bibitem[\protect\citeauthoryear{K.~M.~DeWalt}{K.~M.~DeWalt}{2001}]%
        {b34}
\bibfield{author}{\bibinfo{person}{B.~Wayland K.~M.~DeWalt}.}
  \bibinfo{year}{2001}\natexlab{}.
\newblock \bibinfo{booktitle}{{\em Participant observation: A guide for
  fieldworkers}}.
\newblock \bibinfo{publisher}{AltaMira Press}.
\newblock


\bibitem[\protect\citeauthoryear{Kvale}{Kvale}{1996}]%
        {b35}
\bibfield{author}{\bibinfo{person}{D. Kvale}.} \bibinfo{year}{1996}\natexlab{}.
\newblock \bibinfo{booktitle}{{\em Interviews}}.
\newblock \bibinfo{publisher}{London: SAGE Publications}.
\newblock


\bibitem[\protect\citeauthoryear{Lee}{Lee}{2000}]%
        {b24}
\bibfield{author}{\bibinfo{person}{Y.S. Lee}.} \bibinfo{year}{2000}\natexlab{}.
\newblock \showarticletitle{The sustainability of university-industry research
  collaboration: An empirical assessment}.
\newblock \bibinfo{journal}{{\em The journal of Technology transfer\/}}
  \bibinfo{volume}{25}, \bibinfo{number}{2} (\bibinfo{year}{2000}).
\newblock


\bibitem[\protect\citeauthoryear{Lincoln and Guba}{Lincoln and Guba}{1985}]%
        {Lincoln}
\bibfield{author}{\bibinfo{person}{Y. Lincoln} {and} \bibinfo{person}{E.
  Guba}.} \bibinfo{year}{1985}\natexlab{}.
\newblock \bibinfo{booktitle}{{\em Naturalistic inquiry}}.
\newblock \bibinfo{publisher}{Thousand Oaks Calif.: Sage}.
\newblock


\bibitem[\protect\citeauthoryear{L\o{}berg, Goebel, and Plagemann}{L\o{}berg
  et~al\mbox{.}}{2018}]%
        {b48}
\bibfield{author}{\bibinfo{person}{Fredrik L\o{}berg}, \bibinfo{person}{Vera
  Goebel}, {and} \bibinfo{person}{Thomas Plagemann}.}
  \bibinfo{year}{2018}\natexlab{}.
\newblock \showarticletitle{Quantifying the Signal Quality of Low-Cost
  Respiratory Effort Sensors for Sleep Apnea Monitoring}. In
  \bibinfo{booktitle}{{\em Proceedings of the 3rd International Workshop on
  Multimedia for Personal Health and Health Care}} {\em
  (\bibinfo{series}{HealthMedia'18})}. \bibinfo{pages}{3--11}.
\newblock
\showISBNx{9781450359825}
\showDOI{%
\url{https://doi.org/10.1145/3264996.3264998}}


\bibitem[\protect\citeauthoryear{Marijan}{Marijan}{2015}]%
        {MarijanMP}
\bibfield{author}{\bibinfo{person}{Dusica Marijan}.}
  \bibinfo{year}{2015}\natexlab{}.
\newblock \showarticletitle{Multi-perspective Regression Test Prioritization
  for Time-Constrained Environments}. In \bibinfo{booktitle}{{\em 2015 IEEE
  International Conference on Software Quality, Reliability and Security}}.
  \bibinfo{pages}{157--162}.
\newblock
\showDOI{%
\url{https://doi.org/10.1109/QRS.2015.31}}


\bibitem[\protect\citeauthoryear{{Marijan} and {Gotlieb}}{{Marijan} and
  {Gotlieb}}{2021}]%
        {Marijan}
\bibfield{author}{\bibinfo{person}{D. {Marijan}} {and} \bibinfo{person}{A.
  {Gotlieb}}.} \bibinfo{year}{2021}\natexlab{}.
\newblock \showarticletitle{Industry-Academia Research Collaboration in
  Software Engineering: The Certus Model}.
\newblock \bibinfo{journal}{{\em Information and Software Technology\/}}
  \bibinfo{volume}{132} (\bibinfo{year}{2021}).
\newblock


\bibitem[\protect\citeauthoryear{Marijan, Gotlieb, and Liaaen}{Marijan
  et~al\mbox{.}}{2019}]%
        {Marijan2019ALA}
\bibfield{author}{\bibinfo{person}{Dusica Marijan}, \bibinfo{person}{A.
  Gotlieb}, {and} \bibinfo{person}{Marius Liaaen}.}
  \bibinfo{year}{2019}\natexlab{}.
\newblock \showarticletitle{A learning algorithm for optimizing continuous
  integration development and testing practice}.
\newblock \bibinfo{journal}{{\em Software: Practice and Experience\/}}
  \bibinfo{volume}{49} (\bibinfo{year}{2019}), \bibinfo{pages}{192--213}.
\newblock


\bibitem[\protect\citeauthoryear{Marijan, Gotlieb, and Sen}{Marijan
  et~al\mbox{.}}{2013}]%
        {Marijan2013}
\bibfield{author}{\bibinfo{person}{Dusica Marijan}, \bibinfo{person}{Arnaud
  Gotlieb}, {and} \bibinfo{person}{Sagar Sen}.}
  \bibinfo{year}{2013}\natexlab{}.
\newblock \showarticletitle{Test Case Prioritization for Continuous Regression
  Testing: An Industrial Case Study}. In \bibinfo{booktitle}{{\em 2013 IEEE
  International Conference on Software Maintenance}}.
  \bibinfo{pages}{540--543}.
\newblock
\showDOI{%
\url{https://doi.org/10.1109/ICSM.2013.91}}


\bibitem[\protect\citeauthoryear{Marijan and Liaaen}{Marijan and
  Liaaen}{2016}]%
        {Marijan2016}
\bibfield{author}{\bibinfo{person}{Dusica Marijan} {and}
  \bibinfo{person}{Marius Liaaen}.} \bibinfo{year}{2016}\natexlab{}.
\newblock \showarticletitle{Effect of Time Window on the Performance of
  Continuous Regression Testing}. In \bibinfo{booktitle}{{\em 2016 IEEE
  International Conference on Software Maintenance and Evolution (ICSME)}}.
  \bibinfo{pages}{568--571}.
\newblock
\showDOI{%
\url{https://doi.org/10.1109/ICSME.2016.77}}


\bibitem[\protect\citeauthoryear{Marijan and Liaaen}{Marijan and
  Liaaen}{2017}]%
        {MarijanOB}
\bibfield{author}{\bibinfo{person}{Dusica Marijan} {and}
  \bibinfo{person}{Marius Liaaen}.} \bibinfo{year}{2017}\natexlab{}.
\newblock \showarticletitle{Test Prioritization with Optimally Balanced
  Configuration Coverage}. In \bibinfo{booktitle}{{\em 2017 IEEE 18th
  International Symposium on High Assurance Systems Engineering (HASE)}}.
  \bibinfo{pages}{100--103}.
\newblock
\showDOI{%
\url{https://doi.org/10.1109/HASE.2017.26}}


\bibitem[\protect\citeauthoryear{Marijan and Liaaen}{Marijan and
  Liaaen}{2018}]%
        {MarijanPSR}
\bibfield{author}{\bibinfo{person}{Dusica Marijan} {and}
  \bibinfo{person}{Marius Liaaen}.} \bibinfo{year}{2018}\natexlab{}.
\newblock \showarticletitle{Practical Selective Regression Testing with
  Effective Redundancy in Interleaved Tests}. In \bibinfo{booktitle}{{\em 2018
  IEEE/ACM 40th International Conference on Software Engineering: Software
  Engineering in Practice Track (ICSE-SEIP)}}. \bibinfo{pages}{153--162}.
\newblock


\bibitem[\protect\citeauthoryear{{Marijan}, {Liaaen}, {Gotlieb}, {Sen}, and
  {Ieva}}{{Marijan} et~al\mbox{.}}{2017}]%
        {b52}
\bibfield{author}{\bibinfo{person}{D. {Marijan}}, \bibinfo{person}{M.
  {Liaaen}}, \bibinfo{person}{A. {Gotlieb}}, \bibinfo{person}{S. {Sen}}, {and}
  \bibinfo{person}{C. {Ieva}}.} \bibinfo{year}{2017}\natexlab{}.
\newblock \showarticletitle{TITAN: Test Suite Optimization for Highly
  Configurable Software}. In \bibinfo{booktitle}{{\em 2017 IEEE International
  Conference on Software Testing, Verification and Validation (ICST)}}.
  \bibinfo{pages}{524--531}.
\newblock
\showDOI{%
\url{https://doi.org/10.1109/ICST.2017.60}}


\bibitem[\protect\citeauthoryear{Marijan, Liaaen, and Sen}{Marijan
  et~al\mbox{.}}{2018}]%
        {MarijanDevOps}
\bibfield{author}{\bibinfo{person}{Dusica Marijan}, \bibinfo{person}{Marius
  Liaaen}, {and} \bibinfo{person}{Sagar Sen}.} \bibinfo{year}{2018}\natexlab{}.
\newblock \showarticletitle{DevOps Improvements for Reduced Cycle Times with
  Integrated Test Optimizations for Continuous Integration}. In
  \bibinfo{booktitle}{{\em 2018 IEEE 42nd Annual Computer Software and
  Applications Conference (COMPSAC)}}, Vol.~\bibinfo{volume}{01}.
  \bibinfo{pages}{22--27}.
\newblock
\showDOI{%
\url{https://doi.org/10.1109/COMPSAC.2018.00012}}


\bibitem[\protect\citeauthoryear{Mathiassen}{Mathiassen}{2000}]%
        {b30}
\bibfield{author}{\bibinfo{person}{Lars Mathiassen}.}
  \bibinfo{year}{2000}\natexlab{}.
\newblock \bibinfo{booktitle}{{\em Collaborative Practice Research}}.
\newblock \bibinfo{publisher}{Springer US}, \bibinfo{address}{Boston, MA},
  \bibinfo{pages}{127--148}.
\newblock
\showISBNx{978-0-387-35505-4}
\showDOI{%
\url{https://doi.org/10.1007/978-0-387-35505-4_9}}


\bibitem[\protect\citeauthoryear{Miller and Floricel}{Miller and
  Floricel}{2004}]%
        {b45}
\bibfield{author}{\bibinfo{person}{Roger Miller} {and} \bibinfo{person}{Serghei
  Floricel}.} \bibinfo{year}{2004}\natexlab{}.
\newblock \showarticletitle{Value Creation and Games of Innovation}.
\newblock \bibinfo{journal}{{\em Research-Technology Management\/}}
  \bibinfo{volume}{47}, \bibinfo{number}{6} (\bibinfo{year}{2004}),
  \bibinfo{pages}{25--37}.
\newblock
\showDOI{%
\url{https://doi.org/10.1080/08956308.2004.11671660}}


\bibitem[\protect\citeauthoryear{Mossige, Gotlieb, and Meling}{Mossige
  et~al\mbox{.}}{2017}]%
        {b47}
\bibfield{author}{\bibinfo{person}{Morten Mossige}, \bibinfo{person}{Arnaud
  Gotlieb}, {and} \bibinfo{person}{Hein Meling}.}
  \bibinfo{year}{2017}\natexlab{}.
\newblock \showarticletitle{Deploying Constraint Programming for Testing ABB
  Painting Robots}.
\newblock \bibinfo{journal}{{\em AI Magazine\/}} \bibinfo{volume}{38},
  \bibinfo{number}{2} (\bibinfo{date}{Jul.} \bibinfo{year}{2017}),
  \bibinfo{pages}{94--96}.
\newblock
\showDOI{%
\url{https://doi.org/10.1609/aimag.v38i2.2723}}


\bibitem[\protect\citeauthoryear{Petersen and Engstr\"{o}m}{Petersen and
  Engstr\"{o}m}{2014}]%
        {b41}
\bibfield{author}{\bibinfo{person}{Kai Petersen} {and} \bibinfo{person}{Emelie
  Engstr\"{o}m}.} \bibinfo{year}{2014}\natexlab{}.
\newblock \showarticletitle{Finding Relevant Research Solutions for Practical
  Problems: The Serp Taxonomy Architecture}. In \bibinfo{booktitle}{{\em
  Proceedings of the 2014 International Workshop on Long-Term Industrial
  Collaboration on Software Engineering}} {\em (\bibinfo{series}{WISE '14})}.
  \bibinfo{pages}{13--20}.
\newblock
\showISBNx{9781450330459}
\showDOI{%
\url{https://doi.org/10.1145/2647648.2647650}}


\bibitem[\protect\citeauthoryear{Punter, Krikhaar, and Bril}{Punter
  et~al\mbox{.}}{2006}]%
        {b38}
\bibfield{author}{\bibinfo{person}{Teade Punter}, \bibinfo{person}{Ren\'{e}~L.
  Krikhaar}, {and} \bibinfo{person}{Reinder~J. Bril}.}
  \bibinfo{year}{2006}\natexlab{}.
\newblock \showarticletitle{Sustainable Technology Transfer}. In
  \bibinfo{booktitle}{{\em Proceedings of the 2006 International Workshop on
  Software Technology Transfer in Software Engineering}} {\em
  (\bibinfo{series}{TT '06})}. \bibinfo{pages}{15--18}.
\newblock
\showISBNx{159593412X}
\showDOI{%
\url{https://doi.org/10.1145/1138046.1138052}}


\bibitem[\protect\citeauthoryear{Ruiz-L{\'o}pez, Sen, Jakobsen, Trop{\'e},
  Castle, Hansen, and Nyg{\aa}rd}{Ruiz-L{\'o}pez et~al\mbox{.}}{2019}]%
        {ruiz2019fighthpv}
\bibfield{author}{\bibinfo{person}{Tom{\'a}s Ruiz-L{\'o}pez},
  \bibinfo{person}{Sagar Sen}, \bibinfo{person}{Elisabeth Jakobsen},
  \bibinfo{person}{Ameli Trop{\'e}}, \bibinfo{person}{Philip~E Castle},
  \bibinfo{person}{Bo~Terning Hansen}, {and} \bibinfo{person}{Mari
  Nyg{\aa}rd}.} \bibinfo{year}{2019}\natexlab{}.
\newblock \showarticletitle{FightHPV: design and evaluation of a mobile game to
  raise awareness about human papillomavirus and nudge people to take action
  against cervical cancer}.
\newblock \bibinfo{journal}{{\em JMIR serious games\/}} \bibinfo{volume}{7},
  \bibinfo{number}{2} (\bibinfo{year}{2019}).
\newblock


\bibitem[\protect\citeauthoryear{Runeson and Min\"{o}r}{Runeson and
  Min\"{o}r}{2014}]%
        {b39}
\bibfield{author}{\bibinfo{person}{Per Runeson} {and} \bibinfo{person}{Sten
  Min\"{o}r}.} \bibinfo{year}{2014}\natexlab{}.
\newblock \showarticletitle{The 4+1 View Model of Industry--Academia
  Collaboration}. In \bibinfo{booktitle}{{\em Proceedings of the 2014
  International Workshop on Long-Term Industrial Collaboration on Software
  Engineering}} {\em (\bibinfo{series}{WISE '14})}. \bibinfo{pages}{21--24}.
\newblock
\showISBNx{9781450330459}
\showDOI{%
\url{https://doi.org/10.1145/2647648.2647651}}


\bibitem[\protect\citeauthoryear{Runeson, Min\"{o}r, and Sven\'{e}r}{Runeson
  et~al\mbox{.}}{2014}]%
        {b8}
\bibfield{author}{\bibinfo{person}{Per Runeson}, \bibinfo{person}{Sten
  Min\"{o}r}, {and} \bibinfo{person}{Johan Sven\'{e}r}.}
  \bibinfo{year}{2014}\natexlab{}.
\newblock \showarticletitle{Get the Cogs in Synch: Time Horizon Aspects of
  Industry--Academia Collaboration}. In \bibinfo{booktitle}{{\em Proceedings of
  the 2014 International Workshop on Long-Term Industrial Collaboration on
  Software Engineering}} {\em (\bibinfo{series}{WISE '14})}.
  \bibinfo{pages}{25--28}.
\newblock
\showISBNx{9781450330459}
\showDOI{%
\url{https://doi.org/10.1145/2647648.2647652}}


\bibitem[\protect\citeauthoryear{Schumpeter}{Schumpeter}{1982}]%
        {Schumpeter}
\bibfield{author}{\bibinfo{person}{Joseph~A. Schumpeter}.}
  \bibinfo{year}{1982}\natexlab{}.
\newblock \showarticletitle{The theory of economic development: An inquiry into
  profits, capital, credit, interest, and the business cycle}.
\newblock \bibinfo{journal}{{\em Transaction Publishers\/}}
  \bibinfo{volume}{244} (\bibinfo{year}{1982}), \bibinfo{pages}{1912--1934}.
\newblock


\bibitem[\protect\citeauthoryear{{Selic}}{{Selic}}{2015}]%
        {b5}
\bibfield{author}{\bibinfo{person}{B. {Selic}}.}
  \bibinfo{year}{2015}\natexlab{}.
\newblock \showarticletitle{The Iceberg Effect: On Technology Transfer from
  Research to Practice}. In \bibinfo{booktitle}{{\em 2015 IEEE/ACM 2nd
  International Workshop on Software Engineering Research and Industrial
  Practice}}. \bibinfo{pages}{58--61}.
\newblock
\showDOI{%
\url{https://doi.org/10.1109/SERIP.2015.19}}


\bibitem[\protect\citeauthoryear{Sen, Alesio, Marijan, and Sarkar}{Sen
  et~al\mbox{.}}{2015}]%
        {7302459}
\bibfield{author}{\bibinfo{person}{Sagar Sen}, \bibinfo{person}{Stefano~Di
  Alesio}, \bibinfo{person}{Dusica Marijan}, {and} \bibinfo{person}{Arnab
  Sarkar}.} \bibinfo{year}{2015}\natexlab{}.
\newblock \showarticletitle{Evaluating Reconfiguration Impact in Self-Adaptive
  Systems -- An Approach Based on Combinatorial Interaction Testing}. In
  \bibinfo{booktitle}{{\em 2015 41st Euromicro Conference on Software
  Engineering and Advanced Applications}}. \bibinfo{pages}{250--254}.
\newblock
\showDOI{%
\url{https://doi.org/10.1109/SEAA.2015.72}}


\bibitem[\protect\citeauthoryear{Sen, Bernab{\'{e}}, and Husom}{Sen
  et~al\mbox{.}}{2020}]%
        {DBLP:conf/ijcai/SenBH20}
\bibfield{author}{\bibinfo{person}{Sagar Sen}, \bibinfo{person}{Pierre
  Bernab{\'{e}}}, {and} \bibinfo{person}{Erik Johannes B. L.~G. Husom}.}
  \bibinfo{year}{2020}\natexlab{}.
\newblock \showarticletitle{DeepVentilation: Learning to Predict Physical
  Effort from Breathing}. In \bibinfo{booktitle}{{\em Proceedings of the
  Twenty-Ninth International Joint Conference on Artificial Intelligence,
  {IJCAI} 2020}}, \bibfield{editor}{\bibinfo{person}{Christian Bessiere}}
  (Ed.). \bibinfo{publisher}{ijcai.org}, \bibinfo{pages}{5231--5233}.
\newblock
\showDOI{%
\url{https://doi.org/10.24963/ijcai.2020/753}}


\bibitem[\protect\citeauthoryear{{Sen}, {Marijan}, {Ieva}, {Grime}, and
  {Sander}}{{Sen} et~al\mbox{.}}{2017}]%
        {b44}
\bibfield{author}{\bibinfo{person}{S. {Sen}}, \bibinfo{person}{D. {Marijan}},
  \bibinfo{person}{C. {Ieva}}, \bibinfo{person}{A. {Grime}}, {and}
  \bibinfo{person}{A. {Sander}}.} \bibinfo{year}{2017}\natexlab{}.
\newblock \showarticletitle{Modeling and Verifying Combinatorial Interactions
  to Test Data Intensive Systems: Experience at the Norwegian Customs
  Directorate}.
\newblock \bibinfo{journal}{{\em IEEE Transactions on Reliability\/}}
  \bibinfo{volume}{66}, \bibinfo{number}{1} (\bibinfo{year}{2017}),
  \bibinfo{pages}{3--16}.
\newblock
\showDOI{%
\url{https://doi.org/10.1109/TR.2016.2618121}}


\bibitem[\protect\citeauthoryear{Sharif, Marijan, and Liaaen}{Sharif
  et~al\mbox{.}}{2021}]%
        {Sharif2021DeepOrderDL}
\bibfield{author}{\bibinfo{person}{Aizaz Sharif}, \bibinfo{person}{Dusica
  Marijan}, {and} \bibinfo{person}{Marius Liaaen}.}
  \bibinfo{year}{2021}\natexlab{}.
\newblock \showarticletitle{DeepOrder: Deep Learning for Test Case
  Prioritization in Continuous Integration Testing}.
\newblock \bibinfo{journal}{{\em ArXiv\/}}  \bibinfo{volume}{abs/2110.07443}
  (\bibinfo{year}{2021}).
\newblock


\bibitem[\protect\citeauthoryear{Shneiderman}{Shneiderman}{2018}]%
        {b4}
\bibfield{author}{\bibinfo{person}{Ben Shneiderman}.}
  \bibinfo{year}{2018}\natexlab{}.
\newblock \showarticletitle{Twin-Win Model: A human-centered approach to
  research success}.
\newblock \bibinfo{journal}{{\em Proceedings of the National Academy of
  Sciences\/}} \bibinfo{volume}{115}, \bibinfo{number}{50}
  (\bibinfo{year}{2018}), \bibinfo{pages}{12590--12594}.
\newblock
\showISSN{0027-8424}
\showDOI{%
\url{https://doi.org/10.1073/pnas.1802918115}}
\showeprint{https://www.pnas.org/content/115/50/12590.full.pdf}


\bibitem[\protect\citeauthoryear{Sjoo and Hellstrom}{Sjoo and
  Hellstrom}{2019}]%
        {b16}
\bibfield{author}{\bibinfo{person}{Karolin Sjoo} {and} \bibinfo{person}{Tomas
  Hellstrom}.} \bibinfo{year}{2019}\natexlab{}.
\newblock \showarticletitle{University-industry collaboration: A literature
  review and synthesis}.
\newblock \bibinfo{journal}{{\em Industry and Higher Education\/}}
  \bibinfo{volume}{33}, \bibinfo{number}{4} (\bibinfo{year}{2019}),
  \bibinfo{pages}{275--285}.
\newblock
\showDOI{%
\url{https://doi.org/10.1177/0950422219829697}}
\showeprint{https://doi.org/10.1177/0950422219829697}


\bibitem[\protect\citeauthoryear{{Taylor} and {Bogdan}}{{Taylor} and
  {Bogdan}}{1984}]%
        {b32}
\bibfield{author}{\bibinfo{person}{BS. {Taylor}} {and} \bibinfo{person}{R.
  {Bogdan}}.} \bibinfo{year}{1984}\natexlab{}.
\newblock \bibinfo{booktitle}{{\em Introduction to qualitative research
  methods}}.
\newblock \bibinfo{publisher}{New York: John Wiley and Sons}.
\newblock


\bibitem[\protect\citeauthoryear{{Voas}}{{Voas}}{2004}]%
        {Voas}
\bibfield{author}{\bibinfo{person}{J. {Voas}}.}
  \bibinfo{year}{2004}\natexlab{}.
\newblock \showarticletitle{Software's secret sauce: the "-ilities" [software
  quality]}.
\newblock \bibinfo{journal}{{\em IEEE Software\/}} \bibinfo{volume}{21},
  \bibinfo{number}{6} (\bibinfo{year}{2004}), \bibinfo{pages}{14--15}.
\newblock
\showDOI{%
\url{https://doi.org/10.1109/MS.2004.54}}


\bibitem[\protect\citeauthoryear{{Wohlin}, {Aurum}, {Angelis}, {Phillips},
  {Dittrich}, {Gorschek}, {Grahn}, {Henningsson}, {Kagstrom}, {Low},
  {Rovegard}, {Tomaszewski}, {van Toorn}, and {Winter}}{{Wohlin}
  et~al\mbox{.}}{2012}]%
        {b12}
\bibfield{author}{\bibinfo{person}{C. {Wohlin}}, \bibinfo{person}{A. {Aurum}},
  \bibinfo{person}{L. {Angelis}}, \bibinfo{person}{L. {Phillips}},
  \bibinfo{person}{Y. {Dittrich}}, \bibinfo{person}{T. {Gorschek}},
  \bibinfo{person}{H. {Grahn}}, \bibinfo{person}{K. {Henningsson}},
  \bibinfo{person}{S. {Kagstrom}}, \bibinfo{person}{G. {Low}},
  \bibinfo{person}{P. {Rovegard}}, \bibinfo{person}{P. {Tomaszewski}},
  \bibinfo{person}{C. {van Toorn}}, {and} \bibinfo{person}{J. {Winter}}.}
  \bibinfo{year}{2012}\natexlab{}.
\newblock \showarticletitle{The Success Factors Powering Industry-Academia
  Collaboration}.
\newblock \bibinfo{journal}{{\em IEEE Software\/}} \bibinfo{volume}{29},
  \bibinfo{number}{2} (\bibinfo{year}{2012}), \bibinfo{pages}{67--73}.
\newblock


\bibitem[\protect\citeauthoryear{Y.~Dittrich}{Y.~Dittrich}{2008}]%
        {b28}
\bibfield{author}{\bibinfo{person}{J.~Eriksson C. Hansson O.~Lindeberg
  Y.~Dittrich, K.~Rnkk}.} \bibinfo{year}{2008}\natexlab{}.
\newblock \showarticletitle{Cooperative method development: Combining
  qualitative empirical research with method, technique and process
  improvement}.
\newblock \bibinfo{journal}{{\em Empirical Software Engineering\/}}
  \bibinfo{volume}{13} (\bibinfo{year}{2008}).
\newblock


\end{thebibliography}

\appendix

\end{document}